\documentclass[prl,reprint,superscriptaddress,longbibliography]{revtex4-2}
\usepackage{braket}
\usepackage{amsmath}
\usepackage{graphicx}
\usepackage{natbib}
\usepackage[section]{placeins}
\usepackage{pgfplots,amsfonts}
\usepackage[breaklinks]{hyperref}
\usepackage{amssymb}
\usepackage{amsmath,amsthm,bm}
\usepackage{amsfonts}
\usepackage{cleveref}
\hypersetup{citecolor=red,colorlinks=true,urlcolor=blue}

\DeclareUnicodeCharacter{2212}{-}

\begin{document}

\title{Exact Ansatz of Fermion-Boson Systems for a Quantum Device}

\author{Samuel Warren}
\affiliation{Department of Chemistry and The James Franck Institute, The University of Chicago, Chicago, Illinois 60637, USA}
\author{Yuchen Wang}
\affiliation{Department of Chemistry and The James Franck Institute, The University of Chicago, Chicago, Illinois 60637, USA}
\author{Carlos L. Benavides-Riveros }
\email{cl.benavidesriveros@unitn.it}
\affiliation{Pitaevskii BEC Center, CNR-INO and Dipartimento di Fisica, Università di Trento, I-38123 Trento, Italy}
\author{David A. Mazziotti}
\email{damazz@uchicago.edu}
\affiliation{Department of Chemistry and The James Franck Institute, The University of Chicago, Chicago, Illinois 60637, USA}

\date{Submitted February 16, 2024}

\begin{abstract}
We present an exact ansatz for the eigenstate problem of mixed fer\-mion-boson sys\-tems that can be implemented on quantum devices. Based on a generalization of the electronic contracted Schr\"o\-din\-ger equation (CSE), our approach guides a trial wave function to the ground state of any arbitrary mixed Hamiltonian by directly measuring residuals of the mixed CSE on a quantum device. Unlike density-functional and coupled-cluster theories applied to electron-phonon or electron-photon systems, the accuracy of our approach is not limited by the unknown exchange-correlation functional or the uncontrolled form of the exponential ansatz. To test the performance of the me\-thod, we stu\-dy the Tavis-Cummings model, commonly used in polaritonic quantum chemistry. Our results demonstrate that the CSE is a powerful tool in the development of quantum algorithms for solving general fermion-boson many-body problems.
\end{abstract}

\maketitle

\emph{Introduction}.--- Strong coupling with bosonic particles can drastically change many physical properties of electronic systems. For instance, when coupled with pho\-nons, low-energy electronic excita\-tions are strongly mo\-di\-fied, influencing, as a result, the optical, thermodynamic, and transport properties of solids \cite{10.1093/acprof:oso/9780198507796.001.0001}. This electron-pho\-non coupling is also the source of the effective attractive electronic interaction needed for conventional superconductivity \cite{doi:10.1080/00018735400101213, PhysRev.108.1175}. When coupled with light, emergent hybrid quantum states (known as polaritons) can catalyze or inhibit the reactive paths of chemical reactions  \cite{doi:10.1126/science.abd0336,doi:10.1021/acs.accounts.6b00295,f.ribeiroPolaritonChemistryControlling2018,Thomas2016}. Both electron-phonon and electron-photon couplings lead to many fascinating chemical and technolo\-gi\-cal applications, including spintronics \cite{Mondal2023}, quantum information processing \cite{Jahnke_2015,Ladd2010}, optical control of collective modes in solids \cite{Bloch2022}, catalysis \cite{campos-gonzalez-anguloResonantCatalysisThermally2019, kena-cohenPolaritonChemistryAction2019, duCatalysisDarkStates2022}, solar power \cite{farhatBifacialSchottkyJunctionPlasmonicBased2020}, or low energy lasing \cite{imamog-luNonequilibriumCondensatesLasers1996,dengPolaritonLasingVs2003,kimCoherentPolaritonLaser2016,lengOpticallyPumpedPolaritons2023}.

Given the complexity of these entangled electron-bo\-son systems, it is not surprising that their theoretical description usually relies on semi-empirical model Ha\-mil\-tonians \cite{RevModPhys.89.015003, Walther_2006,PhysRevLett.123.266401,PhysRevX.13.031036}.  However, a more complete understanding of the effects arising from mixed particle coupling requires quan\-titative methods that treat electronic and bosonic modes with equal theoretical rigor \cite{spohn_2004,10.21468/SciPostPhys.9.5.066,C8CS00193F,Ruggen2023}. For instance, to predict reaction pathways in polaritonic catalysis, it is important to port over electronic structure methods to mixed fer\-mion-boson problems \cite{flickInitioOptimizedEffective2018}. Current ab-initio approaches are mainly variants of density functional theory (DFT), and, while lattice dynamics and electron-photon coupling can be accounted for through perturbative \cite{PhysRevLett.68.3603,PhysRevB.94.085415,juraschekHighlyConfinedPhonon2021} or quantum electrodynamics DFT \cite{Ruggenthaler2018,flickInitioOptimizedEffective2018,welakuhTunableNonlinearityEfficient2023}, the unknown form of the corresponding exchange-correlation functionals for the electron-bos\-on interaction limits the accuracy of both approaches \cite{PhysRevB.107.024305,PhysRevX.13.031026,foleyInitioMethodsPolariton2023}. Alternative methods can be found in a new class of coupled cluster (CC) algorithms. Originally developed for electrons \cite{10.1063/1.1727484,Kummel1991}, CC has been extended to electron-phonon \cite{10.1063/5.0033132} and electron-photon \cite{mordovinaPolaritonicCoupledclusterTheory2020,PhysRevX.10.041043,liebenthalEquationofmotionCavityQuantum2022} systems, where the key ingredient is an exponential ansatz believed to span a trial wave function across the most significant regions of the Hilbert space. Additionally, inspired by the prospects of quantum computing, there has been considerable development of CC's unitary form \cite{doi:10.1021/acs.jpclett.1c02659,doi:10.1021/acs.jpclett.3c01935,Denner2023,PhysRevLett.129.066401}. In both cases the expressibility of the approach strongly depends on the quality of the ansatz and, from the quantum computational viewpoint, its Trotterized implementation is not always well defined \cite{doi:10.1021/acs.jctc.9b01083}.

Here we report the development of an alternative approach that gives an exact ansatz for ground and excited states of arbitrary electron-boson systems, overcoming the limitations of both DFT and CC methods. Our approach is based on an extension of the contracted Schr\"odinger equation (CSE)~\cite{mazziottiContractedSchrodingerEquation1998,Colmenero.1993,Nakatsuji.1996,Mazziotti.1999j9j, Mukherjee.2001, Yasuda.2002, Mazziotti.2002, Mazziotti.2020, Smart.2024, cohenHierarchyEquationsReduced1976, PhysRevA.14.41, Valdemoro2007, Mazziotti.20060v3,Mazziotti.2007,Mazziotti.2007k2h,Snyder.2011u3,Gidofalvi.2009,Alcoba.2011,Boyn.2021}, known in the context of reduced density matrix theory for fermionic systems \cite{mazziottiQuantumChemistryWave2006, ch8, mazziottiTwoElectronReducedDensity2012}. Our main result is an exact ansatz that can be implemented directly on quantum devices to find the eigenstates of arbitrary mixed particle Hamiltonians.

The Letter is structured as follows. First, we recap and extend the fermionic CSE to general many-body physics, including boson-fermion systems. We then discuss how the CSE ansatz can inform a quantum algorithm for finding the ground states of mixed particle systems. In the second part of the paper, we demonstrate the effectiveness of the CSE ansatz on the Tavis-Cummings model. Finally, we discuss potential future directions and the implications of our results.

\emph{Theory}.--- Originally derived for fermionic systems, the \emph{Contracted Schr\"{o}dinger Equation} (CSE) reads \cite{mazziottiContractedSchrodingerEquation1998, Colmenero.1993, Nakatsuji.1996, Mazziotti.1999j9j, Mukherjee.2001, Yasuda.2002, Mazziotti.2002, Mazziotti.2020, Smart.2024, cohenHierarchyEquationsReduced1976, PhysRevA.14.41, Valdemoro2007}:
\begin{equation}
\label{cse1}
    \bra{\Psi}\hat{a}_{i_1}^\dagger \hat{a}_{i_2}^\dagger \hat{a}_{k_2}^{} \hat{a}_{k_1}^{} \hat{H}\ket{\Psi} = E \;^2D^{i_1i_2}_{k_1k_2}
\end{equation}
where $\hat{a}_{i}^\dagger$ and $\hat{a}_k$ are fermionic creation and annihilation operators on the $i^{\rm th}$ and $k^{\rm th}$ sites, respectively, and
\begin{equation}
^2D^{i_1i_2}_{k_1k_2} =  \bra{\Psi}\hat{a}_{i_1}^\dagger \hat{a}_{i_2}^\dagger \hat{a}_{k_2}^{} \hat{a}_{k_1}^{} \! \ket{\Psi}
\end{equation}
is the two-body reduced density matrix (RDM). Nakatsuji’s theorem states that the CSE  \eqref{cse1} is satisfied if and only if the corresponding $N$-body preimage of $^2D^{i_1i_2}_{k_1k_2}$ satisﬁes the usual Schr{\"o}dinger equation (SE) \cite{PhysRevA.57.4219, PhysRevA.14.41}.

We now extend the CSE to mixed fermion-boson systems and show that Nakatsuji’s theorem also holds for those systems. Let's first define a general electron-boson density operator
\begin{align}
&\hat\Gamma^{i_1...i_q,j_1...j_s}_{k_1...k_r,l_1...l_t} \nonumber \\  &\qquad = \hat{a}_{i_1}^\dagger...\hat{a}_{i_q}^\dagger \hat{a}_{k_r}^{}...\hat{a}_{k_1}^{} \hat{b}_{j_1}^\dagger...\hat{b}_{j_s}^\dagger \hat{b}_{l_t}^{}...\hat{b}_{l_1}^{} \equiv \hat{\Gamma}^{\bm{i},\bm{j}}_{\bm{k},\bm{l}},
\label{dop}
\end{align}
where $\hat{b}^\dagger_j$ and $\hat{b}_l$ are bosonic creation and annihilation operators. Here $i_m$ and $k_m$ ($j_m$ and $l_m$) run over the different fermionic (bosonic) indices, and we use the compact notation $\bm{i} = (i_1,...,i_q)$, $\bm{k} = (k_1,...,k_r)$, $\bm{j} = (j_1,...,j_s)$ and $\bm{l} = (l_1,...,l_t)$.

The density operator \eqref{dop} allows us to define concisely a general fermion-boson ($fb$) Ha\-mil\-tonian:
\begin{align}
    \hat H_{fb} &=  \sum_{\bm{i}\bm{k},\bm{j}\bm{l}} h^{\bm{i},\bm{j}}_{\bm{k},\bm{l}}\, \hat{\Gamma}^{\bm{i},\bm{j}}_{\bm{k},\bm{l}}. \label{Hamiltonian}
\end{align}
We then multiply the corresponding fermion-boson SE, i.e., $\hat H_{fb} \ket{\Psi} = E \ket{\Psi}$, on the left by $\bra{\Psi}\hat{\Gamma}^{\bm{i},\bm{j}}_{\bm{k},\bm{l}}$ to obtain a \textit{generalized} CSE:
\begin{equation}
    \bra{\Psi}\hat{\Gamma}^{\bm{i},\bm{j}}_{\bm{k},\bm{l}}
    \hat{H}_{fb}\ket{\Psi} = E D^{\bm{i},\bm{j}}_{\bm{k},\bm{l}},
    \label{eq:NCSE}
\end{equation}
where $D^{\bm{i},\bm{j}}_{\bm{k},\bm{l}} = \bra{\Psi}\hat{\Gamma}^{\bm{i},\bm{j}}_{\bm{k},\bm{l}}\ket{\Psi}$ is a \textit{generalized} RDM of the electron-boson system.  Multiplying both sides of the generalized CSE \eqref{eq:NCSE} by the elements of the reduced Hamiltonian matrix $h^{\bm{i},\bm{j}}_{\bm{k},\bm{l}}$ and summing over all indices yields
\begin{equation}
 \sum_{\bm{i}\bm{k},\bm{j}\bm{l}}  h^{\bm{i},\bm{j}}_{\bm{k},\bm{l}}    \bra{\Psi}\hat{\Gamma}^{\bm{i},\bm{j}}_{\bm{k},\bm{l}}
    \hat{H}_{fb}\ket{\Psi} = E \sum_{\bm{i}\bm{k},\bm{j}\bm{l}}  h^{\bm{i},\bm{j}}_{\bm{k},\bm{l}}    D^{\bm{i},\bm{j}}_{\bm{k},\bm{l}}.
    \label{eq:NCSE1}
\end{equation}
The sum on the right-hand side is equal to the energy, making the expression equal to $E^2$. Thus, the equation can be rewritten as the energy variance
\begin{equation}
\bra{\Psi}\hat H_{fb}^2\ket{\Psi}-\bra{\Psi}\hat H_{fb}\ket{\Psi}^2 = 0,
\label{eq:dispersion}
\end{equation}
which, as a stationary condition for the wave function, is equivalent to the SE.  Therefore, the set of solutions to Eq.~\eqref{eq:NCSE} must be the same as the solutions to the electron-boson SE. This derivation shows that the minimal RDM necessary to satisfy both the CSE and SE will have the same degrees of freedom as the corresponding many-body Hamiltonian. This is re\-mi\-niscent of the standard electronic structure problem where a nondegenerate electronic ground-state wavefunction maps to a unique 2-electron RDM, which, as a result, has enough information to build higher-order RDMs and the exact wavefunction \cite{rosina}.

The CSE \eqref{eq:NCSE} can be further decomposed into Hermitian and anti-Hermitian parts:
\begin{multline}
    \bra{\Psi}\hat \Gamma (\hat{H}_{fb}-E))\ket{\Psi} \\
    =\frac{1}{2} \big(\bra{\Psi}[\hat \Gamma ,\hat{H}_{fb}]\ket{\Psi}+\bra{\Psi}\{\hat \Gamma ,\hat{H}_{fb}-E\}\ket{\Psi}\big),
    \label{CSEeb}
\end{multline}
where $[\cdot,\cdot]$ and $\{\cdot,\cdot\}$ are the usual commutator and anticommutator. As described in several prior works for the electronic problem, this decomposition can be used to converge to stationary states either through classical \cite{Mazziotti.20060v3, Mazziotti.2007, Mazziotti.2007k2h, Snyder.2011u3, Gidofalvi.2009, Alcoba.2011, Boyn.2021, PhysRevA.69.012507} or quantum \cite{PhysRevLett.126.070504, Smart.2024, Boyn.2021u94, Smart.2022w8u, Smart.2022, Wang.20232b, wang2023boson} computing methods.

On modern quantum devices, the \textit{Contracted Quantum Eigensolver} (CQE) algorithm measures the total residual of Eq.~\eqref{CSEeb} for trial wave functions. Such a residual can then be used to guide a sequence of trial wave functions toward the ground (or an eigen-) state by iteratively applying a sequence of exponential transformations. The scheme is agnostic to the statistics of the system and has already been applied both for fermions and bosons with significant success \cite{Smart.2024, Boyn.2021u94, Smart.2022w8u, Smart.2022, Wang.20232b, wang2023boson}. Here, we will show that the CQE algorithms also provide a simple methodology for resolving the ground state in mixed fermion-boson systems.

Our scheme is as follows: at iteration $(n+1)$ the wave function results from two separate exponential transformations of the wave function at iteration $(n)$:
\begin{align}
\ket{\Psi^{(n+1)}} = \exp(\eta_B \hat B^{(n)})\exp(\eta_A \hat A^{(n)})   \ket{\Psi^{(n)}},
\end{align}
where
\begin{align}
\hat A^{(n)} = \sum_{\bm{i}\bm{k},\bm{j}\bm{l}} A^{(n)}_{\bm{i}\bm{k},\bm{j}\bm{l}}  \hat{\Gamma}^{\bm{i},\bm{j}}_{\bm{k},\bm{l}}
\end{align}
is an anti-Hermitian operator,
\begin{align}
\hat B^{(n)} = \sum_{\bm{i}\bm{k},\bm{j}\bm{l}} B^{(n)}_{\bm{i}\bm{k},\bm{j}\bm{l}}  \hat{\Gamma}^{\bm{i},\bm{j}}_{\bm{k},\bm{l}}
\end{align}
is a Hermitian one, and $\eta_A$ and $\eta_B$ can be interpreted as learning rates of the algorithm.

Notice that if the unitary operator $\exp(\eta_A \hat A^{(n)})$ is applied to a normalized wave function $\ket{\Psi^{(n)}}$, the total energy of the transformed state is (in leading order of the parameter $\eta_A$):
$\mathcal{E}_{n+1} = \mathcal{E}_{n}  + \eta_A \bra{\Psi^{(n)}} [\hat  H_{fb}, \hat A^{(n)}]  \ket{\Psi^{(n)}} + \mathcal{O}(\eta_A^2)$ where $\mathcal{E}_n = \bra{\Psi^{(n)}}\hat H_{fb}\ket{\Psi^{(n)}}$. As a result, the anti-Hermitian portion of Eq.~\eqref{CSEeb} can be used as a residual to find the optimal operator at each step, and the an\-ti-Hermitian parameters can be updated as follows:
\begin{align}
     A^{(n)} = \bra{\Psi^{(n)}} [\hat\Gamma,\hat  H_{fb}]  \ket{\Psi^{(n)}}.
\end{align}
These parameters generate the unitary $\exp(\eta_A \hat A^{(n)})$. In turn, $\eta_A$ can be selected to minimize the expectation energy of the resulting state $\ket{\Phi^{(n)}} = \exp(\eta_A \hat A^{(n)}) \ket{\Psi^{(n)}}$. Quite remarkably, many well-known canonical trans\-for\-mations of mixed fer\-mion-boson systems are exactly recovered when only the anti-Hermitian part of the ansatz is used. This is the case, for instance, of the Lang-Firsov transformation for the Holstein Hamiltonian \cite{Lang} or the Schrieffer-Wolff transformation for the Anderson model  \cite{PhysRev.149.491}. As already suggested for the electronic case \cite{PhysRevA.69.012507}, this means that the anti-Hermitian part of the CSE converges to the appropriate canonical transformation of the respective problem.

Next, the Hermitian portion of the residual is measured with respect to the updated (normalized) vector $\ket{\Phi^{(n)}}$.
\begin{equation}
     B^{(n)}=
    \bra{\Phi^{(n)}}\{\hat \Gamma ,\hat{H}_{fb}-E^{(n)}\} \ket{\Phi^{(n)}},
\end{equation}
where $E^{(n)}=\bra{\Phi^{(n)}}\hat H_{fb}\ket{\Phi^{(n)}}$. The resulting Hermitian operator $\hat B^{(n)}$ generates the non-unitary $\exp(\eta_B \hat B^{(n)})$ where $\eta_B$ is selected to minimize the energy of $\ket{\Psi^{(n+1)}}$. Implementing non-unitary operators on quantum devices is an active field of research \cite{Hu2020,doi:10.1021/acs.jctc.3c00316} and has resulted in the development of several methods such as quantum imaginary-time evolution \cite{Motta2020,McArdle2019}. Prior implementations of the fermionic CQE have utilized dilation techniques similar to the Sz.-Nagy dilation \cite{Smart.2024}, but the CQE is agnostic to the particular technique used to accomplish non-unitary transformations. Here, we exactly map the non-unitary transformation to a unitary transformation on a classical device, allowing us to focus on the effectiveness of the CSE ansatz for mixed systems.

To study the numerical performance of the algorithm, we use the \textit{Tavis-Cummings model} (TC). This is a prototypical mi\-xed fermion-boson system that attempts to capture the behavior of polaritons in a wide range of coupling regimes~\cite{1443594,tavisExactSolutionMolecule1968,tavisApproximateSolutionsMoleculeRadiationField1969,PhysRevA.105.013719,Castanos_2009}. The model is comprised of $N$ two-level fer\-mio\-nic systems coupled to a bosonic mo\-de, making it ana\-lo\-gous to situations found in pola\-ri\-to\-nic chemistry where molecules are bound in cavities~\cite{kena-cohenPolaritonChemistryAction2019, f.ribeiroPolaritonChemistryControlling2018, geraEffectsDisorderPolaritonic2022,C8CS00193F} or solid-state intersubband devices \cite{PhysRevB.106.224206,PhysRevB.85.045304}. The TC Ha\-mil\-tonian is written as follows:
\begin{align}
\hat H &= \omega_b \hat{b}^\dagger \hat{b} \nonumber \\ & \quad + \sum^N_{i=1} \left[\omega_f \hat{a}^\dagger_{i+} \hat{a}^{}_{i+} + g_c \big( \hat{a}^\dagger_{i+} \hat{a}^{}_{i-} \hat{b} + \hat{a}^\dagger_{i-} \hat{a}^{}_{i+} \hat{b}^\dagger\big)\right],
\label{eq:ham}
\end{align}
with the fermionic subscripts indicating the excited ($+$) and ground ($-$) orbitals in the $i^{\mathrm{th}}$ two-level system, and where $\omega_b$, $\omega_f$, and $g_c$ describe the angular frequency of the bosonic mode, the transition frequency of the fer\-mio\-nic modes, and the coupling between the bosonic mode and the fermionic bath, respectively.

\begin{figure}
\includegraphics[width = \linewidth]{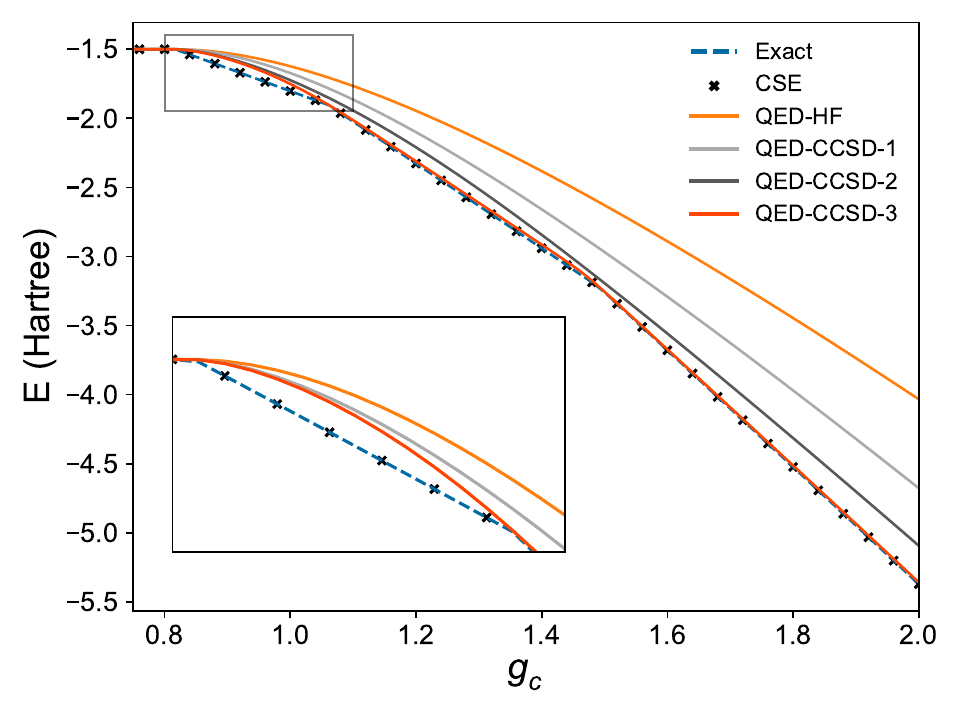}
\caption{CSE and QED-CCSD energies for the 3 fermion Tavis-Cummings model with increasing coupling. The Hamiltonian parameters from Eq.~\eqref{eq:ham} were fixed as $(\omega_b,\omega_f) = (2,0.5)$ while $g_c$ is varied as shown along the $x$-axis. The QED-CCSD-$n$ methods are named according to the convention used in Ref.~\cite{PhysRevX.10.041043}.}
\label{fig:en_ccsd}
\end{figure}
\begin{figure}
\includegraphics[width = \linewidth]{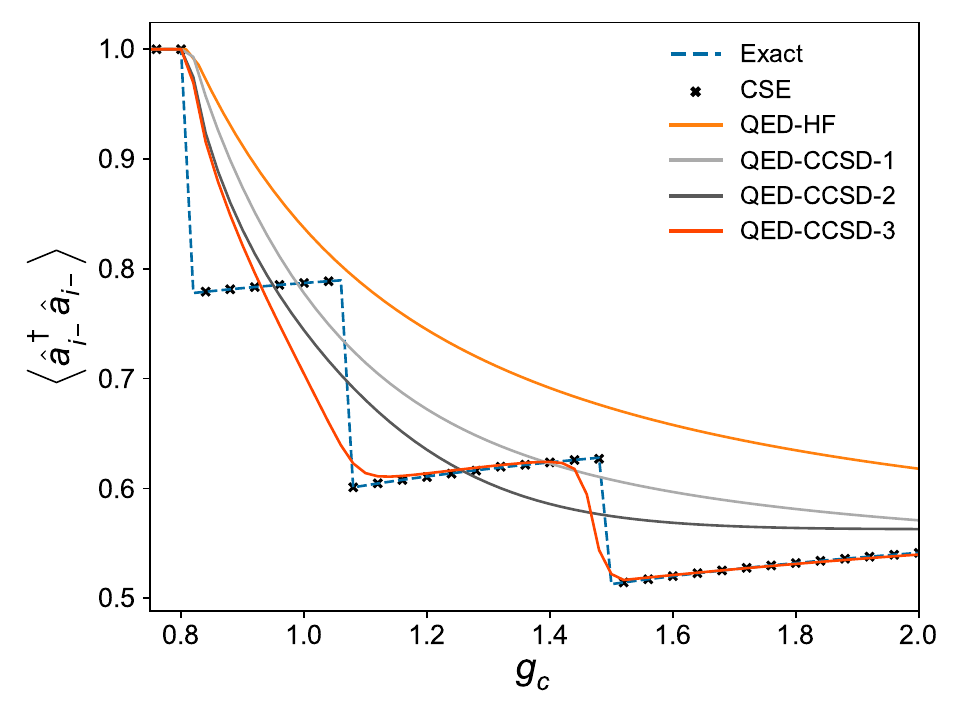}
\caption{CSE and QED-CCSD predicted ground fermionic orbital populations with increasing coupling in the 3-fermion Tavis-Cummings Model.}
\label{fig:pop_ccsd}
\end{figure}
\begin{figure}
\includegraphics[width = \linewidth]{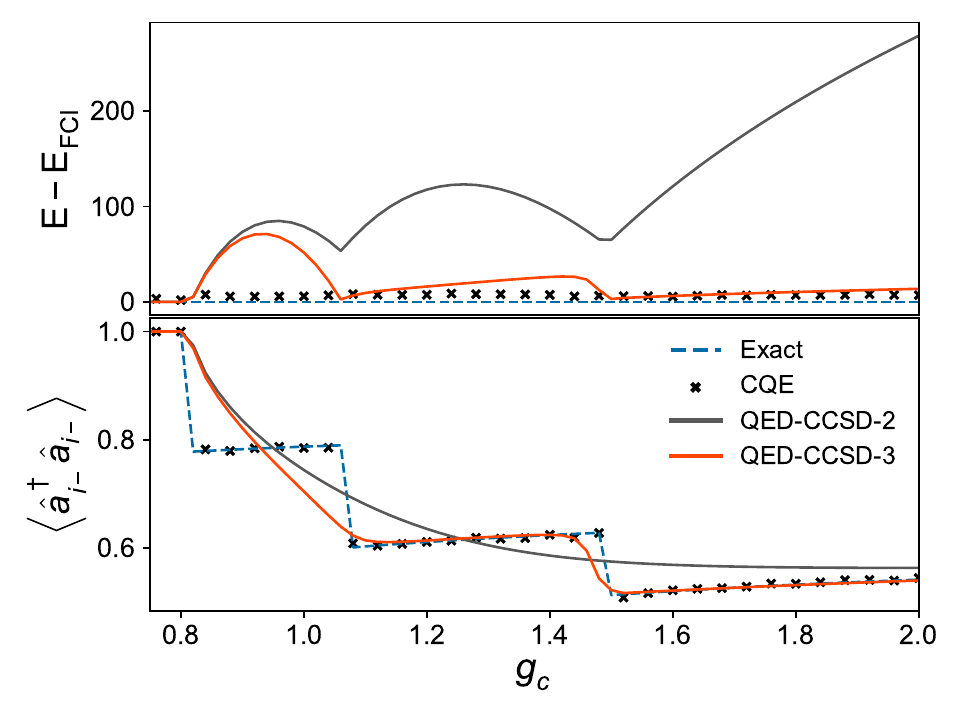}
\caption{CQE and QED-CCSD ground-state energy and fermionic orbital population with increasing coupling. The $y$-axis of the energy shows the absolute error in mHartree.}
\label{fig:en_cqe}
\end{figure}

\emph{Results}.---
We first compare the CQE ground-state results to those obtained from the mixed fermion-boson quantum electrodynamics coupled cluster (QED-CC) methods for the TC model.
Those are named such that QED-CC(SD)-$n$ refers to coupled-clus\-ter with single and double fermionic cluster terms combined with bo\-so\-nic and mixed cluster terms containing up to $n$ bo\-so\-nic creation operators \cite{PhysRevX.10.041043}.

Fig.~\ref{fig:en_ccsd} depicts the ground-state energies predicted by various methods for the 3-site TC model as the coupling strength increases, entering well into the ultra-strong re\-gi\-me. The QED-CC methods capture the general trend of the energies but do not align with any section of the exact energy curve, including the high-level QED-CCSD-3, which is the most accurate QED-CCSD method possible in this coupling regime due to the maximum bosonic population of 3. This suggests that QED-CCSD-3 would begin to fail again in the stronger coupling limit when the maximum bosonic population increases to 4 or larger. Additionally, despite being the best CCSD method in this region, QED-CCSD-3 is unable to predict the transition in the weak coupling limit seen in the zoomed-in section of the figure due to the omission of the triple cluster operator. The CSE, however, is able to accurately reproduce the exact energy regardless of coupling strength. Notice that this accuracy is achieved while only ever measuring the residual with the same density operators that appear in the TC Hamiltonian, while QED-CC must include all possible excitations to be accurate. Therefore, while the number of bosonic and mixed cluster terms remains fixed for the CQE, for the QED-CC methods it will grow with the coupling strength or, equivalently, the maximum boson population. 

Fig.~\ref{fig:pop_ccsd} demonstrates how QED-CC methods fail to capture both the quantitative and qualitative properties of the TC model. The plot shows the predicted population of the lower energy level in any of the fermionic two-level subsystems, which, due to symmetry, are the same. The QED-CC methods up to QED-CCSD-2 approximate the discontinuous population changes with successively better least squares errors, but fail to capture any of the stair-stepping behavior caused by actual level crossings. QED-CCSD-3 smoothes out the final two stair steps, but completely misses the first, as could be predicted from its failure in the prior figure. The CSE method, however, exactly recovers the populations, despite needing to resolve nearly degenerate states at the level crossings. The failure of CC to resolve level crossings ~\cite{kohnCanCoupledclusterTheory2007} and the success of the CSE highlight the importance of the particular combination of Hermitian and anti-Hermitian contributions present in the CSE ansatz.

Finally, Fig.~\ref{fig:en_cqe} presents the results from the CQE algorithm and the quantum analogs of the QED-CCSD methods run on an ideal quantum device simulator. The CQE algorithm reproduces the energies with an average error of 7~mhartree with a standard deviation of 1~mhartree, indicating that the error is uniform regardless of the distance from the level crossings. This is in contrast to the CC methods where the error exhibits significant fluctuations. The CQE recovers the ground orbital populations near the degeneracies despite not exactly recovering the energy. These results demonstrate that despite sampling error, the CQE is able to outperform some of the most popular classical algorithms for mixed-particle systems.

\emph{Conclusions}.---   Based on a generalization of the electronic CSE, we present an exact ansatz for mixed quantum fermion-boson systems. Our numerical results on the Tavis-Cumming model demonstrate the power of the CSE ansatz. While CC methods require constantly increasing the number of terms in the exponential to be accurate in different coupling regimes, the CSE is always exact and the required number of terms exactly matches the density operators present in the Hamiltonian. Additionally, the CQE demonstrates how the CSE can be applied on a quantum device, and can outperform cutting-edge classical algorithms. This Letter leaves many avenues for future works, whether it be the development of classical computing algorithms for mixed particle systems that leverage the CSE or the application of the CQE on real quantum devices. Furthermore, since the ansatz can be equally used for any eigenstate \cite{benavidesriveros2023quantum}, the computation of exact excited states in polaritonic quantum chemistry is also a promising future direction.

\hspace{0.05cm}

\begin{acknowledgments}
C.L.B.-R.~gratefully thanks Daniele De Bernardis for insightful discussions and acknowledges financial support from the European Union’s Horizon Europe Re\-search and Innovation program  un\-der the Marie Skło\-dowska-Curie Grant Agreement n°101065295--RDMFTforbosons. Views and opinions expressed are however those of the author only and do not necessarily reflect those of the European Union or the European Research Executive Agency. D.A.M gratefully acknowledges the U.S. Department of Energy, Office of Basic Energy Sciences, Grant DE-SC0019215, and the U.S. National Science Foundation Grant No. CHE-2155082.
\end{acknowledgments}

\bibliography{mixqcse}

\begin{thebibliography}{94}%
\makeatletter
\providecommand \@ifxundefined [1]{%
 \@ifx{#1\undefined}
}%
\providecommand \@ifnum [1]{%
 \ifnum #1\expandafter \@firstoftwo
 \else \expandafter \@secondoftwo
 \fi
}%
\providecommand \@ifx [1]{%
 \ifx #1\expandafter \@firstoftwo
 \else \expandafter \@secondoftwo
 \fi
}%
\providecommand \natexlab [1]{#1}%
\providecommand \enquote  [1]{``#1''}%
\providecommand \bibnamefont  [1]{#1}%
\providecommand \bibfnamefont [1]{#1}%
\providecommand \citenamefont [1]{#1}%
\providecommand \href@noop [0]{\@secondoftwo}%
\providecommand \href [0]{\begingroup \@sanitize@url \@href}%
\providecommand \@href[1]{\@@startlink{#1}\@@href}%
\providecommand \@@href[1]{\endgroup#1\@@endlink}%
\providecommand \@sanitize@url [0]{\catcode `\\12\catcode `\$12\catcode
  `\&12\catcode `\#12\catcode `\^12\catcode `\_12\catcode `\%12\relax}%
\providecommand \@@startlink[1]{}%
\providecommand \@@endlink[0]{}%
\providecommand \url  [0]{\begingroup\@sanitize@url \@url }%
\providecommand \@url [1]{\endgroup\@href {#1}{\urlprefix }}%
\providecommand \urlprefix  [0]{URL }%
\providecommand \Eprint [0]{\href }%
\providecommand \doibase [0]{https://doi.org/}%
\providecommand \selectlanguage [0]{\@gobble}%
\providecommand \bibinfo  [0]{\@secondoftwo}%
\providecommand \bibfield  [0]{\@secondoftwo}%
\providecommand \translation [1]{[#1]}%
\providecommand \BibitemOpen [0]{}%
\providecommand \bibitemStop [0]{}%
\providecommand \bibitemNoStop [0]{.\EOS\space}%
\providecommand \EOS [0]{\spacefactor3000\relax}%
\providecommand \BibitemShut  [1]{\csname bibitem#1\endcsname}%
\let\auto@bib@innerbib\@empty
\bibitem [{\citenamefont
  {Ziman}(2001)}]{10.1093/acprof:oso/9780198507796.001.0001}%
  \BibitemOpen
  \bibfield  {author} {\bibinfo {author} {\bibfnamefont {J.}~\bibnamefont
  {Ziman}},\ }\href {https://doi.org/10.1093/acprof:oso/9780198507796.001.0001}
  {\emph {\bibinfo {title} {{Electrons and Phonons: The Theory of Transport
  Phenomena in Solids}}}}\ (\bibinfo  {publisher} {Oxford University Press},\
  \bibinfo {year} {2001})\BibitemShut {NoStop}%
\bibitem [{\citenamefont {Fröhlich}(1954)}]{doi:10.1080/00018735400101213}%
  \BibitemOpen
  \bibfield  {author} {\bibinfo {author} {\bibfnamefont {H.}~\bibnamefont
  {Fröhlich}},\ }\bibfield  {title} {\bibinfo {title} {Electrons in lattice
  fields},\ }\href {https://doi.org/10.1080/00018735400101213} {\bibfield
  {journal} {\bibinfo  {journal} {Adv. Phys.}\ }\textbf {\bibinfo {volume}
  {3}},\ \bibinfo {pages} {325} (\bibinfo {year} {1954})}\BibitemShut {NoStop}%
\bibitem [{\citenamefont {Bardeen}\ \emph {et~al.}(1957)\citenamefont
  {Bardeen}, \citenamefont {Cooper},\ and\ \citenamefont
  {Schrieffer}}]{PhysRev.108.1175}%
  \BibitemOpen
  \bibfield  {author} {\bibinfo {author} {\bibfnamefont {J.}~\bibnamefont
  {Bardeen}}, \bibinfo {author} {\bibfnamefont {L.~N.}\ \bibnamefont
  {Cooper}},\ and\ \bibinfo {author} {\bibfnamefont {J.~R.}\ \bibnamefont
  {Schrieffer}},\ }\bibfield  {title} {\bibinfo {title} {Theory of
  superconductivity},\ }\href {https://doi.org/10.1103/PhysRev.108.1175}
  {\bibfield  {journal} {\bibinfo  {journal} {Phys. Rev.}\ }\textbf {\bibinfo
  {volume} {108}},\ \bibinfo {pages} {1175} (\bibinfo {year}
  {1957})}\BibitemShut {NoStop}%
\bibitem [{\citenamefont {Garcia-Vidal}\ \emph {et~al.}(2021)\citenamefont
  {Garcia-Vidal}, \citenamefont {Ciuti},\ and\ \citenamefont
  {Ebbesen}}]{doi:10.1126/science.abd0336}%
  \BibitemOpen
  \bibfield  {author} {\bibinfo {author} {\bibfnamefont {F.~J.}\ \bibnamefont
  {Garcia-Vidal}}, \bibinfo {author} {\bibfnamefont {C.}~\bibnamefont
  {Ciuti}},\ and\ \bibinfo {author} {\bibfnamefont {T.~W.}\ \bibnamefont
  {Ebbesen}},\ }\bibfield  {title} {\bibinfo {title} {Manipulating matter by
  strong coupling to vacuum fields},\ }\href
  {https://doi.org/10.1126/science.abd0336} {\bibfield  {journal} {\bibinfo
  {journal} {Science}\ }\textbf {\bibinfo {volume} {373}},\ \bibinfo {pages}
  {eabd0336} (\bibinfo {year} {2021})}\BibitemShut {NoStop}%
\bibitem [{\citenamefont {Ebbesen}(2016)}]{doi:10.1021/acs.accounts.6b00295}%
  \BibitemOpen
  \bibfield  {author} {\bibinfo {author} {\bibfnamefont {T.~W.}\ \bibnamefont
  {Ebbesen}},\ }\bibfield  {title} {\bibinfo {title} {Hybrid light–matter
  states in a molecular and material science perspective},\ }\href
  {https://doi.org/10.1021/acs.accounts.6b00295} {\bibfield  {journal}
  {\bibinfo  {journal} {Acc. Chem. Res}\ }\textbf {\bibinfo {volume} {49}},\
  \bibinfo {pages} {2403} (\bibinfo {year} {2016})}\BibitemShut {NoStop}%
\bibitem [{\citenamefont {Ribeiro}\ \emph {et~al.}(2018)\citenamefont
  {Ribeiro}, \citenamefont {{A.~Mart{\'i}nez-Mart{\'i}nez}}, \citenamefont
  {Du}, \citenamefont {{Campos-Gonzalez-Angulo}},\ and\ \citenamefont
  {{Yuen-Zhou}}}]{f.ribeiroPolaritonChemistryControlling2018}%
  \BibitemOpen
  \bibfield  {author} {\bibinfo {author} {\bibfnamefont {R.}~\bibnamefont
  {Ribeiro}}, \bibinfo {author} {\bibfnamefont {L.}~\bibnamefont
  {{A.~Mart{\'i}nez-Mart{\'i}nez}}}, \bibinfo {author} {\bibfnamefont
  {M.}~\bibnamefont {Du}}, \bibinfo {author} {\bibfnamefont {J.}~\bibnamefont
  {{Campos-Gonzalez-Angulo}}},\ and\ \bibinfo {author} {\bibfnamefont
  {J.}~\bibnamefont {{Yuen-Zhou}}},\ }\bibfield  {title} {\bibinfo {title}
  {{Polariton Chemistry: Controlling Molecular Dynamics with Optical
  Cavities}},\ }\href {https://doi.org/10.1039/C8SC01043A} {\bibfield
  {journal} {\bibinfo  {journal} {Chem. Sci.}\ }\textbf {\bibinfo {volume}
  {9}},\ \bibinfo {pages} {6325} (\bibinfo {year} {2018})}\BibitemShut
  {NoStop}%
\bibitem [{\citenamefont {Thomas}\ \emph {et~al.}(2016)\citenamefont {Thomas},
  \citenamefont {George}, \citenamefont {Shalabney}, \citenamefont {Dryzhakov},
  \citenamefont {Varma}, \citenamefont {Moran}, \citenamefont {Chervy},
  \citenamefont {Zhong}, \citenamefont {Devaux}, \citenamefont {Genet},
  \citenamefont {Hutchison},\ and\ \citenamefont {Ebbesen}}]{Thomas2016}%
  \BibitemOpen
  \bibfield  {author} {\bibinfo {author} {\bibfnamefont {A.}~\bibnamefont
  {Thomas}}, \bibinfo {author} {\bibfnamefont {J.}~\bibnamefont {George}},
  \bibinfo {author} {\bibfnamefont {A.}~\bibnamefont {Shalabney}}, \bibinfo
  {author} {\bibfnamefont {M.}~\bibnamefont {Dryzhakov}}, \bibinfo {author}
  {\bibfnamefont {S.}~\bibnamefont {Varma}}, \bibinfo {author} {\bibfnamefont
  {J.}~\bibnamefont {Moran}}, \bibinfo {author} {\bibfnamefont
  {T.}~\bibnamefont {Chervy}}, \bibinfo {author} {\bibfnamefont
  {X.}~\bibnamefont {Zhong}}, \bibinfo {author} {\bibfnamefont
  {E.}~\bibnamefont {Devaux}}, \bibinfo {author} {\bibfnamefont
  {C.}~\bibnamefont {Genet}}, \bibinfo {author} {\bibfnamefont {J.~A.}\
  \bibnamefont {Hutchison}},\ and\ \bibinfo {author} {\bibfnamefont
  {T.}~\bibnamefont {Ebbesen}},\ }\bibfield  {title} {\bibinfo {title}
  {{Ground-State Chemical Reactivity under Vibrational Coupling to the Vacuum
  Electromagnetic Field}},\ }\href
  {https://doi.org/https://doi.org/10.1002/anie.201605504} {\bibfield
  {journal} {\bibinfo  {journal} {Angew. Chem. Int. Ed.}\ }\textbf {\bibinfo
  {volume} {55}},\ \bibinfo {pages} {11462} (\bibinfo {year}
  {2016})}\BibitemShut {NoStop}%
\bibitem [{\citenamefont {Mondal}\ and\ \citenamefont
  {Lunghi}(2023)}]{Mondal2023}%
  \BibitemOpen
  \bibfield  {author} {\bibinfo {author} {\bibfnamefont {S.}~\bibnamefont
  {Mondal}}\ and\ \bibinfo {author} {\bibfnamefont {A.}~\bibnamefont
  {Lunghi}},\ }\bibfield  {title} {\bibinfo {title} {Spin-phonon decoherence in
  solid-state paramagnetic defects from first principles},\ }\href
  {https://doi.org/10.1038/s41524-023-01082-9} {\bibfield  {journal} {\bibinfo
  {journal} {npj Comput. Mater.}\ }\textbf {\bibinfo {volume} {9}},\ \bibinfo
  {pages} {120} (\bibinfo {year} {2023})}\BibitemShut {NoStop}%
\bibitem [{\citenamefont {Jahnke}\ \emph {et~al.}(2015)\citenamefont {Jahnke},
  \citenamefont {Sipahigil}, \citenamefont {Binder}, \citenamefont {Doherty},
  \citenamefont {Metsch}, \citenamefont {Rogers}, \citenamefont {Manson},
  \citenamefont {Lukin},\ and\ \citenamefont {Jelezko}}]{Jahnke_2015}%
  \BibitemOpen
  \bibfield  {author} {\bibinfo {author} {\bibfnamefont {K.}~\bibnamefont
  {Jahnke}}, \bibinfo {author} {\bibfnamefont {A.}~\bibnamefont {Sipahigil}},
  \bibinfo {author} {\bibfnamefont {J.}~\bibnamefont {Binder}}, \bibinfo
  {author} {\bibfnamefont {M.}~\bibnamefont {Doherty}}, \bibinfo {author}
  {\bibfnamefont {M.}~\bibnamefont {Metsch}}, \bibinfo {author} {\bibfnamefont
  {L.}~\bibnamefont {Rogers}}, \bibinfo {author} {\bibfnamefont
  {N.}~\bibnamefont {Manson}}, \bibinfo {author} {\bibfnamefont
  {M.}~\bibnamefont {Lukin}},\ and\ \bibinfo {author} {\bibfnamefont
  {F.}~\bibnamefont {Jelezko}},\ }\bibfield  {title} {\bibinfo {title}
  {Electron–phonon processes of the silicon-va\-can\-cy centre in diamond},\
  }\href {https://doi.org/10.1088/1367-2630/17/4/043011} {\bibfield  {journal}
  {\bibinfo  {journal} {New J. Phys.}\ }\textbf {\bibinfo {volume} {17}},\
  \bibinfo {pages} {043011} (\bibinfo {year} {2015})}\BibitemShut {NoStop}%
\bibitem [{\citenamefont {Ladd}\ \emph {et~al.}(2010)\citenamefont {Ladd},
  \citenamefont {Jelezko}, \citenamefont {Laflamme}, \citenamefont {Nakamura},
  \citenamefont {Monroe},\ and\ \citenamefont {O'Brien}}]{Ladd2010}%
  \BibitemOpen
  \bibfield  {author} {\bibinfo {author} {\bibfnamefont {T.~D.}\ \bibnamefont
  {Ladd}}, \bibinfo {author} {\bibfnamefont {F.}~\bibnamefont {Jelezko}},
  \bibinfo {author} {\bibfnamefont {R.}~\bibnamefont {Laflamme}}, \bibinfo
  {author} {\bibfnamefont {Y.}~\bibnamefont {Nakamura}}, \bibinfo {author}
  {\bibfnamefont {C.}~\bibnamefont {Monroe}},\ and\ \bibinfo {author}
  {\bibfnamefont {J.~L.}\ \bibnamefont {O'Brien}},\ }\bibfield  {title}
  {\bibinfo {title} {Quantum computers},\ }\href
  {https://doi.org/10.1038/nature08812} {\bibfield  {journal} {\bibinfo
  {journal} {Nature}\ }\textbf {\bibinfo {volume} {464}},\ \bibinfo {pages}
  {45} (\bibinfo {year} {2010})}\BibitemShut {NoStop}%
\bibitem [{\citenamefont {Bloch}\ \emph {et~al.}(2022)\citenamefont {Bloch},
  \citenamefont {Cavalleri}, \citenamefont {Galitski}, \citenamefont {Hafezi},\
  and\ \citenamefont {Rubio}}]{Bloch2022}%
  \BibitemOpen
  \bibfield  {author} {\bibinfo {author} {\bibfnamefont {J.}~\bibnamefont
  {Bloch}}, \bibinfo {author} {\bibfnamefont {A.}~\bibnamefont {Cavalleri}},
  \bibinfo {author} {\bibfnamefont {V.}~\bibnamefont {Galitski}}, \bibinfo
  {author} {\bibfnamefont {M.}~\bibnamefont {Hafezi}},\ and\ \bibinfo {author}
  {\bibfnamefont {A.}~\bibnamefont {Rubio}},\ }\bibfield  {title} {\bibinfo
  {title} {Strongly correlated electron--photon systems},\ }\href
  {https://doi.org/10.1038/s41586-022-04726-w} {\bibfield  {journal} {\bibinfo
  {journal} {Nature}\ }\textbf {\bibinfo {volume} {606}},\ \bibinfo {pages}
  {41} (\bibinfo {year} {2022})}\BibitemShut {NoStop}%
\bibitem [{\citenamefont {{Campos-Gonzalez-Angulo}}\ \emph
  {et~al.}(2019)\citenamefont {{Campos-Gonzalez-Angulo}}, \citenamefont
  {Ribeiro},\ and\ \citenamefont
  {{Yuen-Zhou}}}]{campos-gonzalez-anguloResonantCatalysisThermally2019}%
  \BibitemOpen
  \bibfield  {author} {\bibinfo {author} {\bibfnamefont {J.}~\bibnamefont
  {{Campos-Gonzalez-Angulo}}}, \bibinfo {author} {\bibfnamefont
  {R.}~\bibnamefont {Ribeiro}},\ and\ \bibinfo {author} {\bibfnamefont
  {J.}~\bibnamefont {{Yuen-Zhou}}},\ }\bibfield  {title} {\bibinfo {title}
  {{Resonant Catalysis of Thermally Activated Chemical Reactions with
  Vibrational Polaritons}},\ }\href
  {https://doi.org/10.1038/s41467-019-12636-1} {\bibfield  {journal} {\bibinfo
  {journal} {Nat. Commun.}\ }\textbf {\bibinfo {volume} {10}},\ \bibinfo
  {pages} {4685} (\bibinfo {year} {2019})}\BibitemShut {NoStop}%
\bibitem [{\citenamefont {{K{\'e}na-Cohen}}\ and\ \citenamefont
  {{Yuen-Zhou}}(2019)}]{kena-cohenPolaritonChemistryAction2019}%
  \BibitemOpen
  \bibfield  {author} {\bibinfo {author} {\bibfnamefont {S.}~\bibnamefont
  {{K{\'e}na-Cohen}}}\ and\ \bibinfo {author} {\bibfnamefont {J.}~\bibnamefont
  {{Yuen-Zhou}}},\ }\bibfield  {title} {\bibinfo {title} {Polariton
  {{Chemistry}}: {{Action}} in the {{Dark}}},\ }\href
  {https://doi.org/10.1021/acscentsci.9b00219} {\bibfield  {journal} {\bibinfo
  {journal} {ACS Cent. Sci.}\ }\textbf {\bibinfo {volume} {5}},\ \bibinfo
  {pages} {386} (\bibinfo {year} {2019})}\BibitemShut {NoStop}%
\bibitem [{\citenamefont {Du}\ and\ \citenamefont
  {{Yuen-Zhou}}(2022)}]{duCatalysisDarkStates2022}%
  \BibitemOpen
  \bibfield  {author} {\bibinfo {author} {\bibfnamefont {M.}~\bibnamefont
  {Du}}\ and\ \bibinfo {author} {\bibfnamefont {J.}~\bibnamefont
  {{Yuen-Zhou}}},\ }\bibfield  {title} {\bibinfo {title} {Catalysis by {{Dark
  States}} in {{Vibropolaritonic Chemistry}}},\ }\href
  {https://doi.org/10.1103/PhysRevLett.128.096001} {\bibfield  {journal}
  {\bibinfo  {journal} {Phys. Rev. Lett.}\ }\textbf {\bibinfo {volume} {128}},\
  \bibinfo {pages} {096001} (\bibinfo {year} {2022})}\BibitemShut {NoStop}%
\bibitem [{\citenamefont {Farhat}\ \emph {et~al.}(2020)\citenamefont {Farhat},
  \citenamefont {Baloch}, \citenamefont {Rashkeev}, \citenamefont {Tabet},
  \citenamefont {Kais},\ and\ \citenamefont
  {Alharbi}}]{farhatBifacialSchottkyJunctionPlasmonicBased2020}%
  \BibitemOpen
  \bibfield  {author} {\bibinfo {author} {\bibfnamefont {M.}~\bibnamefont
  {Farhat}}, \bibinfo {author} {\bibfnamefont {A.}~\bibnamefont {Baloch}},
  \bibinfo {author} {\bibfnamefont {S.}~\bibnamefont {Rashkeev}}, \bibinfo
  {author} {\bibfnamefont {N.}~\bibnamefont {Tabet}}, \bibinfo {author}
  {\bibfnamefont {S.}~\bibnamefont {Kais}},\ and\ \bibinfo {author}
  {\bibfnamefont {F.}~\bibnamefont {Alharbi}},\ }\bibfield  {title} {\bibinfo
  {title} {Bifacial {{Schottky-Junction Plasmonic-Based Solar Cell}}},\ }\href
  {https://doi.org/10.1002/ente.201901280} {\bibfield  {journal} {\bibinfo
  {journal} {Energy Technol.}\ }\textbf {\bibinfo {volume} {8}},\ \bibinfo
  {pages} {1901280} (\bibinfo {year} {2020})}\BibitemShut {NoStop}%
\bibitem [{\citenamefont {Imamoglu}\ \emph {et~al.}(1996)\citenamefont
  {Imamoglu}, \citenamefont {Ram}, \citenamefont {Pau},\ and\ \citenamefont
  {Yamamoto}}]{imamog-luNonequilibriumCondensatesLasers1996}%
  \BibitemOpen
  \bibfield  {author} {\bibinfo {author} {\bibfnamefont {A.}~\bibnamefont
  {Imamoglu}}, \bibinfo {author} {\bibfnamefont {R.}~\bibnamefont {Ram}},
  \bibinfo {author} {\bibfnamefont {S.}~\bibnamefont {Pau}},\ and\ \bibinfo
  {author} {\bibfnamefont {Y.}~\bibnamefont {Yamamoto}},\ }\bibfield  {title}
  {\bibinfo {title} {Nonequilibrium condensates and lasers without inversion:
  {{Exciton-polariton}} lasers},\ }\href
  {https://doi.org/10.1103/PhysRevA.53.4250} {\bibfield  {journal} {\bibinfo
  {journal} {Phys. Rev. A}\ }\textbf {\bibinfo {volume} {53}},\ \bibinfo
  {pages} {4250} (\bibinfo {year} {1996})}\BibitemShut {NoStop}%
\bibitem [{\citenamefont {Deng}\ \emph {et~al.}(2003)\citenamefont {Deng},
  \citenamefont {Weihs}, \citenamefont {Snoke}, \citenamefont {Bloch},\ and\
  \citenamefont {Yamamoto}}]{dengPolaritonLasingVs2003}%
  \BibitemOpen
  \bibfield  {author} {\bibinfo {author} {\bibfnamefont {H.}~\bibnamefont
  {Deng}}, \bibinfo {author} {\bibfnamefont {G.}~\bibnamefont {Weihs}},
  \bibinfo {author} {\bibfnamefont {D.}~\bibnamefont {Snoke}}, \bibinfo
  {author} {\bibfnamefont {J.}~\bibnamefont {Bloch}},\ and\ \bibinfo {author}
  {\bibfnamefont {Y.}~\bibnamefont {Yamamoto}},\ }\bibfield  {title} {\bibinfo
  {title} {{Polariton Lasing vs. Photon Lasing in a Semiconductor
  Microcavity}},\ }\href {https://doi.org/10.1073/pnas.2634328100} {\bibfield
  {journal} {\bibinfo  {journal} {Proc. Natl. Acad. Sci. U.S.A.}\ }\textbf
  {\bibinfo {volume} {100}},\ \bibinfo {pages} {15318} (\bibinfo {year}
  {2003})}\BibitemShut {NoStop}%
\bibitem [{\citenamefont {Kim}\ \emph {et~al.}(2016)\citenamefont {Kim},
  \citenamefont {Zhang}, \citenamefont {Wang}, \citenamefont {Fischer},
  \citenamefont {Brodbeck}, \citenamefont {Kamp}, \citenamefont {Schneider},
  \citenamefont {H{\"o}fling},\ and\ \citenamefont
  {Deng}}]{kimCoherentPolaritonLaser2016}%
  \BibitemOpen
  \bibfield  {author} {\bibinfo {author} {\bibfnamefont {S.}~\bibnamefont
  {Kim}}, \bibinfo {author} {\bibfnamefont {B.}~\bibnamefont {Zhang}}, \bibinfo
  {author} {\bibfnamefont {Z.}~\bibnamefont {Wang}}, \bibinfo {author}
  {\bibfnamefont {J.}~\bibnamefont {Fischer}}, \bibinfo {author} {\bibfnamefont
  {S.}~\bibnamefont {Brodbeck}}, \bibinfo {author} {\bibfnamefont
  {M.}~\bibnamefont {Kamp}}, \bibinfo {author} {\bibfnamefont {C.}~\bibnamefont
  {Schneider}}, \bibinfo {author} {\bibfnamefont {S.}~\bibnamefont
  {H{\"o}fling}},\ and\ \bibinfo {author} {\bibfnamefont {H.}~\bibnamefont
  {Deng}},\ }\bibfield  {title} {\bibinfo {title} {Coherent {{Polariton
  Laser}}},\ }\href {https://doi.org/10.1103/PhysRevX.6.011026} {\bibfield
  {journal} {\bibinfo  {journal} {Phys. Rev. X}\ }\textbf {\bibinfo {volume}
  {6}},\ \bibinfo {pages} {011026} (\bibinfo {year} {2016})}\BibitemShut
  {NoStop}%
\bibitem [{\citenamefont {Leng}\ \emph {et~al.}(2023)\citenamefont {Leng},
  \citenamefont {Wu}, \citenamefont {Dini}, \citenamefont {Liu}, \citenamefont
  {Hu}, \citenamefont {Tang}, \citenamefont {Liew}, \citenamefont {Sun},
  \citenamefont {Su},\ and\ \citenamefont
  {Xiong}}]{lengOpticallyPumpedPolaritons2023}%
  \BibitemOpen
  \bibfield  {author} {\bibinfo {author} {\bibfnamefont {M.}~\bibnamefont
  {Leng}}, \bibinfo {author} {\bibfnamefont {J.}~\bibnamefont {Wu}}, \bibinfo
  {author} {\bibfnamefont {K.}~\bibnamefont {Dini}}, \bibinfo {author}
  {\bibfnamefont {J.}~\bibnamefont {Liu}}, \bibinfo {author} {\bibfnamefont
  {Z.}~\bibnamefont {Hu}}, \bibinfo {author} {\bibfnamefont {J.}~\bibnamefont
  {Tang}}, \bibinfo {author} {\bibfnamefont {T.~C.~H.}\ \bibnamefont {Liew}},
  \bibinfo {author} {\bibfnamefont {H.}~\bibnamefont {Sun}}, \bibinfo {author}
  {\bibfnamefont {R.}~\bibnamefont {Su}},\ and\ \bibinfo {author}
  {\bibfnamefont {Q.}~\bibnamefont {Xiong}},\ }\bibfield  {title} {\bibinfo
  {title} {Optically {{Pumped Polaritons}} in {{Perovskite Light-Emitting
  Diodes}}},\ }\href {https://doi.org/10.1021/acsphotonics.2c01999} {\bibfield
  {journal} {\bibinfo  {journal} {ACS Photonics}\ }\textbf {\bibinfo {volume}
  {10}},\ \bibinfo {pages} {1349} (\bibinfo {year} {2023})}\BibitemShut
  {NoStop}%
\bibitem [{\citenamefont {Giustino}(2017)}]{RevModPhys.89.015003}%
  \BibitemOpen
  \bibfield  {author} {\bibinfo {author} {\bibfnamefont {F.}~\bibnamefont
  {Giustino}},\ }\bibfield  {title} {\bibinfo {title} {Electron-phonon
  interactions from first principles},\ }\href
  {https://doi.org/10.1103/RevModPhys.89.015003} {\bibfield  {journal}
  {\bibinfo  {journal} {Rev. Mod. Phys.}\ }\textbf {\bibinfo {volume} {89}},\
  \bibinfo {pages} {015003} (\bibinfo {year} {2017})}\BibitemShut {NoStop}%
\bibitem [{\citenamefont {Walther}\ \emph {et~al.}(2006)\citenamefont
  {Walther}, \citenamefont {Varcoe}, \citenamefont {Englert},\ and\
  \citenamefont {Becker}}]{Walther_2006}%
  \BibitemOpen
  \bibfield  {author} {\bibinfo {author} {\bibfnamefont {H.}~\bibnamefont
  {Walther}}, \bibinfo {author} {\bibfnamefont {B.}~\bibnamefont {Varcoe}},
  \bibinfo {author} {\bibfnamefont {B.-G.}\ \bibnamefont {Englert}},\ and\
  \bibinfo {author} {\bibfnamefont {T.}~\bibnamefont {Becker}},\ }\bibfield
  {title} {\bibinfo {title} {Cavity quantum electrodynamics},\ }\href
  {https://doi.org/10.1088/0034-4885/69/5/R02} {\bibfield  {journal} {\bibinfo
  {journal} {Rep. Prog. Phys.}\ }\textbf {\bibinfo {volume} {69}},\ \bibinfo
  {pages} {1325} (\bibinfo {year} {2006})}\BibitemShut {NoStop}%
\bibitem [{\citenamefont {Levinsen}\ \emph {et~al.}(2019)\citenamefont
  {Levinsen}, \citenamefont {Marchetti}, \citenamefont {Keeling},\ and\
  \citenamefont {Parish}}]{PhysRevLett.123.266401}%
  \BibitemOpen
  \bibfield  {author} {\bibinfo {author} {\bibfnamefont {J.}~\bibnamefont
  {Levinsen}}, \bibinfo {author} {\bibfnamefont {F.}~\bibnamefont {Marchetti}},
  \bibinfo {author} {\bibfnamefont {J.}~\bibnamefont {Keeling}},\ and\ \bibinfo
  {author} {\bibfnamefont {M.}~\bibnamefont {Parish}},\ }\bibfield  {title}
  {\bibinfo {title} {{Spectroscopic Signatures of Quantum Many-Body
  Correlations in Polariton Microcavities}},\ }\href
  {https://doi.org/10.1103/PhysRevLett.123.266401} {\bibfield  {journal}
  {\bibinfo  {journal} {Phys. Rev. Lett.}\ }\textbf {\bibinfo {volume} {123}},\
  \bibinfo {pages} {266401} (\bibinfo {year} {2019})}\BibitemShut {NoStop}%
\bibitem [{\citenamefont {Tan}\ \emph {et~al.}(2023)\citenamefont {Tan},
  \citenamefont {Diessel}, \citenamefont {Popert}, \citenamefont {Schmidt},
  \citenamefont {Imamoglu},\ and\ \citenamefont {Kroner}}]{PhysRevX.13.031036}%
  \BibitemOpen
  \bibfield  {author} {\bibinfo {author} {\bibfnamefont {L.}~\bibnamefont
  {Tan}}, \bibinfo {author} {\bibfnamefont {O.}~\bibnamefont {Diessel}},
  \bibinfo {author} {\bibfnamefont {A.}~\bibnamefont {Popert}}, \bibinfo
  {author} {\bibfnamefont {R.}~\bibnamefont {Schmidt}}, \bibinfo {author}
  {\bibfnamefont {A.}~\bibnamefont {Imamoglu}},\ and\ \bibinfo {author}
  {\bibfnamefont {M.}~\bibnamefont {Kroner}},\ }\bibfield  {title} {\bibinfo
  {title} {{Bose Polaron Interactions in a Cavity-Coupled Monolayer
  Semiconductor}},\ }\href {https://doi.org/10.1103/PhysRevX.13.031036}
  {\bibfield  {journal} {\bibinfo  {journal} {Phys. Rev. X}\ }\textbf {\bibinfo
  {volume} {13}},\ \bibinfo {pages} {031036} (\bibinfo {year}
  {2023})}\BibitemShut {NoStop}%
\bibitem [{\citenamefont {Spohn}(2004)}]{spohn_2004}%
  \BibitemOpen
  \bibfield  {author} {\bibinfo {author} {\bibfnamefont {H.}~\bibnamefont
  {Spohn}},\ }\href {https://doi.org/10.1017/CBO9780511535178} {\emph {\bibinfo
  {title} {Dynamics of Charged Particles and their Radiation Field}}}\
  (\bibinfo  {publisher} {Cambridge University Press},\ \bibinfo {year}
  {2004})\BibitemShut {NoStop}%
\bibitem [{\citenamefont {Schuler}\ \emph {et~al.}(2020)\citenamefont
  {Schuler}, \citenamefont {Bernardis}, \citenamefont {Läuchli},\ and\
  \citenamefont {Rabl}}]{10.21468/SciPostPhys.9.5.066}%
  \BibitemOpen
  \bibfield  {author} {\bibinfo {author} {\bibfnamefont {M.}~\bibnamefont
  {Schuler}}, \bibinfo {author} {\bibfnamefont {D.~D.}\ \bibnamefont
  {Bernardis}}, \bibinfo {author} {\bibfnamefont {A.~M.}\ \bibnamefont
  {Läuchli}},\ and\ \bibinfo {author} {\bibfnamefont {P.}~\bibnamefont
  {Rabl}},\ }\bibfield  {title} {\bibinfo {title} {{The vacua of dipolar cavity
  quantum electrodynamics}},\ }\href
  {https://doi.org/10.21468/SciPostPhys.9.5.066} {\bibfield  {journal}
  {\bibinfo  {journal} {SciPost Phys.}\ }\textbf {\bibinfo {volume} {9}},\
  \bibinfo {pages} {066} (\bibinfo {year} {2020})}\BibitemShut {NoStop}%
\bibitem [{\citenamefont {Hertzog}\ \emph {et~al.}(2019)\citenamefont
  {Hertzog}, \citenamefont {Wang}, \citenamefont {Mony},\ and\ \citenamefont
  {Börjesson}}]{C8CS00193F}%
  \BibitemOpen
  \bibfield  {author} {\bibinfo {author} {\bibfnamefont {M.}~\bibnamefont
  {Hertzog}}, \bibinfo {author} {\bibfnamefont {M.}~\bibnamefont {Wang}},
  \bibinfo {author} {\bibfnamefont {J.}~\bibnamefont {Mony}},\ and\ \bibinfo
  {author} {\bibfnamefont {K.}~\bibnamefont {Börjesson}},\ }\bibfield  {title}
  {\bibinfo {title} {Strong light–matter interactions: a new direction within
  che\-mis\-try},\ }\href {https://doi.org/10.1039/C8CS00193F} {\bibfield
  {journal} {\bibinfo  {journal} {Chem. Soc. Rev.}\ }\textbf {\bibinfo {volume}
  {48}},\ \bibinfo {pages} {937} (\bibinfo {year} {2019})}\BibitemShut
  {NoStop}%
\bibitem [{\citenamefont {Ruggenthaler}\ \emph {et~al.}(2023)\citenamefont
  {Ruggenthaler}, \citenamefont {Sidler},\ and\ \citenamefont
  {Rubio}}]{Ruggen2023}%
  \BibitemOpen
  \bibfield  {author} {\bibinfo {author} {\bibfnamefont {M.}~\bibnamefont
  {Ruggenthaler}}, \bibinfo {author} {\bibfnamefont {D.}~\bibnamefont
  {Sidler}},\ and\ \bibinfo {author} {\bibfnamefont {A.}~\bibnamefont
  {Rubio}},\ }\bibfield  {title} {\bibinfo {title} {{Understanding Polaritonic
  Chemistry from Ab Initio Quantum Electrodynamics}},\ }\href
  {https://doi.org/10.1021/acs.chemrev.2c00788} {\bibfield  {journal} {\bibinfo
   {journal} {Chem. Rev.}\ }\textbf {\bibinfo {volume} {123}},\ \bibinfo
  {pages} {11191} (\bibinfo {year} {2023})}\BibitemShut {NoStop}%
\bibitem [{\citenamefont {Flick}\ \emph {et~al.}(2018)\citenamefont {Flick},
  \citenamefont {Sch{\"a}fer}, \citenamefont {Ruggenthaler}, \citenamefont
  {Appel},\ and\ \citenamefont {Rubio}}]{flickInitioOptimizedEffective2018}%
  \BibitemOpen
  \bibfield  {author} {\bibinfo {author} {\bibfnamefont {J.}~\bibnamefont
  {Flick}}, \bibinfo {author} {\bibfnamefont {C.}~\bibnamefont {Sch{\"a}fer}},
  \bibinfo {author} {\bibfnamefont {M.}~\bibnamefont {Ruggenthaler}}, \bibinfo
  {author} {\bibfnamefont {H.}~\bibnamefont {Appel}},\ and\ \bibinfo {author}
  {\bibfnamefont {A.}~\bibnamefont {Rubio}},\ }\bibfield  {title} {\bibinfo
  {title} {Ab {{Initio Optimized Effective Potentials}} for {{Real Molecules}}
  in {{Optical Cavities}}: {{Photon Contributions}} to the {{Molecular Ground
  State}}},\ }\href {https://doi.org/10.1021/acsphotonics.7b01279} {\bibfield
  {journal} {\bibinfo  {journal} {ACS Photonics}\ }\textbf {\bibinfo {volume}
  {5}},\ \bibinfo {pages} {992} (\bibinfo {year} {2018})}\BibitemShut {NoStop}%
\bibitem [{\citenamefont {Gonze}\ \emph {et~al.}(1992)\citenamefont {Gonze},
  \citenamefont {Allan},\ and\ \citenamefont {Teter}}]{PhysRevLett.68.3603}%
  \BibitemOpen
  \bibfield  {author} {\bibinfo {author} {\bibfnamefont {X.}~\bibnamefont
  {Gonze}}, \bibinfo {author} {\bibfnamefont {D.~C.}\ \bibnamefont {Allan}},\
  and\ \bibinfo {author} {\bibfnamefont {M.~P.}\ \bibnamefont {Teter}},\
  }\bibfield  {title} {\bibinfo {title} {Dielectric tensor, effective charges,
  and phonons in \ensuremath{\alpha}-quartz by variational density-functional
  perturbation theory},\ }\href {https://doi.org/10.1103/PhysRevLett.68.3603}
  {\bibfield  {journal} {\bibinfo  {journal} {Phys. Rev. Lett.}\ }\textbf
  {\bibinfo {volume} {68}},\ \bibinfo {pages} {3603} (\bibinfo {year}
  {1992})}\BibitemShut {NoStop}%
\bibitem [{\citenamefont {Sohier}\ \emph {et~al.}(2016)\citenamefont {Sohier},
  \citenamefont {Calandra},\ and\ \citenamefont {Mauri}}]{PhysRevB.94.085415}%
  \BibitemOpen
  \bibfield  {author} {\bibinfo {author} {\bibfnamefont {T.}~\bibnamefont
  {Sohier}}, \bibinfo {author} {\bibfnamefont {M.}~\bibnamefont {Calandra}},\
  and\ \bibinfo {author} {\bibfnamefont {F.}~\bibnamefont {Mauri}},\ }\bibfield
   {title} {\bibinfo {title} {{Two-di\-men\-sio\-nal Fr\"ohlich interaction in
  transition-metal dichalcogenide monolayers: Theoretical modeling and
  first-prin\-ci\-ples calculations}},\ }\href
  {https://doi.org/10.1103/PhysRevB.94.085415} {\bibfield  {journal} {\bibinfo
  {journal} {Phys. Rev. B}\ }\textbf {\bibinfo {volume} {94}},\ \bibinfo
  {pages} {085415} (\bibinfo {year} {2016})}\BibitemShut {NoStop}%
\bibitem [{\citenamefont {Juraschek}\ and\ \citenamefont
  {Narang}(2021)}]{juraschekHighlyConfinedPhonon2021}%
  \BibitemOpen
  \bibfield  {author} {\bibinfo {author} {\bibfnamefont {D.~M.}\ \bibnamefont
  {Juraschek}}\ and\ \bibinfo {author} {\bibfnamefont {P.}~\bibnamefont
  {Narang}},\ }\bibfield  {title} {\bibinfo {title} {Highly {{Confined Phonon
  Polaritons}} in {{Monolayers}} of {{Perovskite Oxides}}},\ }\href
  {https://doi.org/10.1021/acs.nanolett.1c01002} {\bibfield  {journal}
  {\bibinfo  {journal} {Nano Letters}\ }\textbf {\bibinfo {volume} {21}},\
  \bibinfo {pages} {5098} (\bibinfo {year} {2021})}\BibitemShut {NoStop}%
\bibitem [{\citenamefont {Ruggenthaler}\ \emph {et~al.}(2018)\citenamefont
  {Ruggenthaler}, \citenamefont {Tancogne-Dejean}, \citenamefont {Flick},
  \citenamefont {Appel},\ and\ \citenamefont {Rubio}}]{Ruggenthaler2018}%
  \BibitemOpen
  \bibfield  {author} {\bibinfo {author} {\bibfnamefont {M.}~\bibnamefont
  {Ruggenthaler}}, \bibinfo {author} {\bibfnamefont {N.}~\bibnamefont
  {Tancogne-Dejean}}, \bibinfo {author} {\bibfnamefont {J.}~\bibnamefont
  {Flick}}, \bibinfo {author} {\bibfnamefont {H.}~\bibnamefont {Appel}},\ and\
  \bibinfo {author} {\bibfnamefont {A.}~\bibnamefont {Rubio}},\ }\bibfield
  {title} {\bibinfo {title} {From a quantum-electrodynamical light--matter
  description to novel spectroscopies},\ }\href
  {https://doi.org/10.1038/s41570-018-0118} {\bibfield  {journal} {\bibinfo
  {journal} {Nat. Rev. Chem.}\ }\textbf {\bibinfo {volume} {2}},\ \bibinfo
  {pages} {0118} (\bibinfo {year} {2018})}\BibitemShut {NoStop}%
\bibitem [{\citenamefont {Welakuh}\ and\ \citenamefont
  {Narang}(2023)}]{welakuhTunableNonlinearityEfficient2023}%
  \BibitemOpen
  \bibfield  {author} {\bibinfo {author} {\bibfnamefont {D.~M.}\ \bibnamefont
  {Welakuh}}\ and\ \bibinfo {author} {\bibfnamefont {P.}~\bibnamefont
  {Narang}},\ }\bibfield  {title} {\bibinfo {title} {Tunable {{Nonlinearity}}
  and {{Efficient Harmonic Generation}} from a {{Strongly Coupled
  Light}}{\textendash}{{Matter System}}},\ }\href
  {https://doi.org/10.1021/acsphotonics.2c00966} {\bibfield  {journal}
  {\bibinfo  {journal} {ACS Photonics}\ }\textbf {\bibinfo {volume} {10}},\
  \bibinfo {pages} {383} (\bibinfo {year} {2023})}\BibitemShut {NoStop}%
\bibitem [{\citenamefont {Marini}(2023)}]{PhysRevB.107.024305}%
  \BibitemOpen
  \bibfield  {author} {\bibinfo {author} {\bibfnamefont {A.}~\bibnamefont
  {Marini}},\ }\bibfield  {title} {\bibinfo {title} {Equilibrium and
  out-of-equilibrium realistic phonon self-energy free from overscreening},\
  }\href {https://doi.org/10.1103/PhysRevB.107.024305} {\bibfield  {journal}
  {\bibinfo  {journal} {Phys. Rev. B}\ }\textbf {\bibinfo {volume} {107}},\
  \bibinfo {pages} {024305} (\bibinfo {year} {2023})}\BibitemShut {NoStop}%
\bibitem [{\citenamefont {Stefanucci}\ \emph {et~al.}(2023)\citenamefont
  {Stefanucci}, \citenamefont {van Leeuwen},\ and\ \citenamefont
  {Perfetto}}]{PhysRevX.13.031026}%
  \BibitemOpen
  \bibfield  {author} {\bibinfo {author} {\bibfnamefont {G.}~\bibnamefont
  {Stefanucci}}, \bibinfo {author} {\bibfnamefont {R.}~\bibnamefont {van
  Leeuwen}},\ and\ \bibinfo {author} {\bibfnamefont {E.}~\bibnamefont
  {Perfetto}},\ }\bibfield  {title} {\bibinfo {title} {{In and
  Out-of-Equilibrium Ab Initio Theory of Electrons and Phonons}},\ }\href
  {https://doi.org/10.1103/PhysRevX.13.031026} {\bibfield  {journal} {\bibinfo
  {journal} {Phys. Rev. X}\ }\textbf {\bibinfo {volume} {13}},\ \bibinfo
  {pages} {031026} (\bibinfo {year} {2023})}\BibitemShut {NoStop}%
\bibitem [{\citenamefont {Foley~IV}\ \emph {et~al.}(2023)\citenamefont
  {Foley~IV}, \citenamefont {McTague},\ and\ \citenamefont
  {DePrince~III}}]{foleyInitioMethodsPolariton2023}%
  \BibitemOpen
  \bibfield  {author} {\bibinfo {author} {\bibfnamefont {J.}~\bibnamefont
  {Foley~IV}}, \bibinfo {author} {\bibfnamefont {J.}~\bibnamefont {McTague}},\
  and\ \bibinfo {author} {\bibfnamefont {A.}~\bibnamefont {DePrince~III}},\
  }\bibfield  {title} {\bibinfo {title} {{Ab Initio Methods for Polariton
  Chemistry}},\ }\href {https://doi.org/10.1063/5.0167243} {\bibfield
  {journal} {\bibinfo  {journal} {Chem. Phys. Rev.}\ }\textbf {\bibinfo
  {volume} {4}},\ \bibinfo {pages} {041301} (\bibinfo {year}
  {2023})}\BibitemShut {NoStop}%
\bibitem [{\citenamefont {Čížek}(1966)}]{10.1063/1.1727484}%
  \BibitemOpen
  \bibfield  {author} {\bibinfo {author} {\bibfnamefont {J.}~\bibnamefont
  {Čížek}},\ }\bibfield  {title} {\bibinfo {title} {{On the Correlation
  Problem in Atomic and Molecular Systems. Calculation of Wavefunction
  Components in Ursell‐Type Expansion Using Quantum‐Field Theoretical
  Methods}},\ }\href {https://doi.org/10.1063/1.1727484} {\bibfield  {journal}
  {\bibinfo  {journal} {J. Chem. Phys.}\ }\textbf {\bibinfo {volume} {45}},\
  \bibinfo {pages} {4256} (\bibinfo {year} {1966})}\BibitemShut {NoStop}%
\bibitem [{\citenamefont {K{\"u}mmel}(1991)}]{Kummel1991}%
  \BibitemOpen
  \bibfield  {author} {\bibinfo {author} {\bibfnamefont {H.}~\bibnamefont
  {K{\"u}mmel}},\ }\bibfield  {title} {\bibinfo {title} {{Origins of the
  Coupled Cluster Method}},\ }\href {https://doi.org/10.1007/BF01119615}
  {\bibfield  {journal} {\bibinfo  {journal} {Theoret. Chim. Acta}\ }\textbf
  {\bibinfo {volume} {80}},\ \bibinfo {pages} {81} (\bibinfo {year}
  {1991})}\BibitemShut {NoStop}%
\bibitem [{\citenamefont {White}\ \emph {et~al.}(2020)\citenamefont {White},
  \citenamefont {Gao}, \citenamefont {Minnich},\ and\ \citenamefont
  {Chan}}]{10.1063/5.0033132}%
  \BibitemOpen
  \bibfield  {author} {\bibinfo {author} {\bibfnamefont {A.}~\bibnamefont
  {White}}, \bibinfo {author} {\bibfnamefont {Y.}~\bibnamefont {Gao}}, \bibinfo
  {author} {\bibfnamefont {A.~J.}\ \bibnamefont {Minnich}},\ and\ \bibinfo
  {author} {\bibfnamefont {G.}~\bibnamefont {Chan}},\ }\bibfield  {title}
  {\bibinfo {title} {{A cou\-pled cluster framework for electrons and
  phonons}},\ }\href {https://doi.org/10.1063/5.0033132} {\bibfield  {journal}
  {\bibinfo  {journal} {J. Chem. Phys.}\ }\textbf {\bibinfo {volume} {153}},\
  \bibinfo {pages} {224112} (\bibinfo {year} {2020})}\BibitemShut {NoStop}%
\bibitem [{\citenamefont {Mordovina}\ \emph {et~al.}(2020)\citenamefont
  {Mordovina}, \citenamefont {Bungey}, \citenamefont {Appel}, \citenamefont
  {Knowles}, \citenamefont {Rubio},\ and\ \citenamefont
  {Manby}}]{mordovinaPolaritonicCoupledclusterTheory2020}%
  \BibitemOpen
  \bibfield  {author} {\bibinfo {author} {\bibfnamefont {U.}~\bibnamefont
  {Mordovina}}, \bibinfo {author} {\bibfnamefont {C.}~\bibnamefont {Bungey}},
  \bibinfo {author} {\bibfnamefont {H.}~\bibnamefont {Appel}}, \bibinfo
  {author} {\bibfnamefont {P.~J.}\ \bibnamefont {Knowles}}, \bibinfo {author}
  {\bibfnamefont {A.}~\bibnamefont {Rubio}},\ and\ \bibinfo {author}
  {\bibfnamefont {F.~R.}\ \bibnamefont {Manby}},\ }\bibfield  {title} {\bibinfo
  {title} {Polaritonic coupled-cluster theory},\ }\href
  {https://doi.org/10.1103/PhysRevResearch.2.023262} {\bibfield  {journal}
  {\bibinfo  {journal} {Phys. Rev. Res.}\ }\textbf {\bibinfo {volume} {2}},\
  \bibinfo {pages} {023262} (\bibinfo {year} {2020})}\BibitemShut {NoStop}%
\bibitem [{\citenamefont {Haugland}\ \emph {et~al.}(2020)\citenamefont
  {Haugland}, \citenamefont {Ronca}, \citenamefont {Kj\o{}nstad}, \citenamefont
  {Rubio},\ and\ \citenamefont {Koch}}]{PhysRevX.10.041043}%
  \BibitemOpen
  \bibfield  {author} {\bibinfo {author} {\bibfnamefont {T.~S.}\ \bibnamefont
  {Haugland}}, \bibinfo {author} {\bibfnamefont {E.}~\bibnamefont {Ronca}},
  \bibinfo {author} {\bibfnamefont {E.~F.}\ \bibnamefont {Kj\o{}nstad}},
  \bibinfo {author} {\bibfnamefont {A.}~\bibnamefont {Rubio}},\ and\ \bibinfo
  {author} {\bibfnamefont {H.}~\bibnamefont {Koch}},\ }\bibfield  {title}
  {\bibinfo {title} {{Coupled Cluster Theory for Molecular Polaritons: Changing
  Ground and Excited States}},\ }\href
  {https://doi.org/10.1103/PhysRevX.10.041043} {\bibfield  {journal} {\bibinfo
  {journal} {Phys. Rev. X}\ }\textbf {\bibinfo {volume} {10}},\ \bibinfo
  {pages} {041043} (\bibinfo {year} {2020})}\BibitemShut {NoStop}%
\bibitem [{\citenamefont {Liebenthal}\ \emph {et~al.}(2022)\citenamefont
  {Liebenthal}, \citenamefont {Vu},\ and\ \citenamefont
  {DePrince~III}}]{liebenthalEquationofmotionCavityQuantum2022}%
  \BibitemOpen
  \bibfield  {author} {\bibinfo {author} {\bibfnamefont {M.}~\bibnamefont
  {Liebenthal}}, \bibinfo {author} {\bibfnamefont {N.}~\bibnamefont {Vu}},\
  and\ \bibinfo {author} {\bibfnamefont {A.}~\bibnamefont {DePrince~III}},\
  }\bibfield  {title} {\bibinfo {title} {Equation-of-motion cavity quantum
  electrodynamics coupled-cluster theory for electron attachment},\ }\href
  {https://doi.org/10.1063/5.0078795} {\bibfield  {journal} {\bibinfo
  {journal} {J. Chem. Phys.}\ }\textbf {\bibinfo {volume} {156}},\ \bibinfo
  {pages} {054105} (\bibinfo {year} {2022})}\BibitemShut {NoStop}%
\bibitem [{\citenamefont {Pavošević}\ and\ \citenamefont
  {Flick}(2021)}]{doi:10.1021/acs.jpclett.1c02659}%
  \BibitemOpen
  \bibfield  {author} {\bibinfo {author} {\bibfnamefont {F.}~\bibnamefont
  {Pavošević}}\ and\ \bibinfo {author} {\bibfnamefont {J.}~\bibnamefont
  {Flick}},\ }\bibfield  {title} {\bibinfo {title} {{Polaritonic Unitary
  Coupled Cluster for Quantum Computations}},\ }\href
  {https://doi.org/10.1021/acs.jpclett.1c02659} {\bibfield  {journal} {\bibinfo
   {journal} {J. Phys. Chem. Lett.}\ }\textbf {\bibinfo {volume} {12}},\
  \bibinfo {pages} {9100} (\bibinfo {year} {2021})}\BibitemShut {NoStop}%
\bibitem [{\citenamefont {Pavošević}\ \emph {et~al.}(2023)\citenamefont
  {Pavošević}, \citenamefont {Tavernelli},\ and\ \citenamefont
  {Rubio}}]{doi:10.1021/acs.jpclett.3c01935}%
  \BibitemOpen
  \bibfield  {author} {\bibinfo {author} {\bibfnamefont {F.}~\bibnamefont
  {Pavošević}}, \bibinfo {author} {\bibfnamefont {I.}~\bibnamefont
  {Tavernelli}},\ and\ \bibinfo {author} {\bibfnamefont {A.}~\bibnamefont
  {Rubio}},\ }\bibfield  {title} {\bibinfo {title} {{Spin-Flip Unitary Coupled
  Cluster Method: Toward Accurate Description of Strong Electron Correlation on
  Quantum Computers}},\ }\href {https://doi.org/10.1021/acs.jpclett.3c01935}
  {\bibfield  {journal} {\bibinfo  {journal} {J. Phys. Chem. Lett.}\ }\textbf
  {\bibinfo {volume} {14}},\ \bibinfo {pages} {7876} (\bibinfo {year}
  {2023})}\BibitemShut {NoStop}%
\bibitem [{\citenamefont {Denner}\ \emph {et~al.}(2023)\citenamefont {Denner},
  \citenamefont {Miessen}, \citenamefont {Yan}, \citenamefont {Tavernelli},
  \citenamefont {Neupert}, \citenamefont {Demler},\ and\ \citenamefont
  {Wang}}]{Denner2023}%
  \BibitemOpen
  \bibfield  {author} {\bibinfo {author} {\bibfnamefont {M.}~\bibnamefont
  {Denner}}, \bibinfo {author} {\bibfnamefont {A.}~\bibnamefont {Miessen}},
  \bibinfo {author} {\bibfnamefont {H.}~\bibnamefont {Yan}}, \bibinfo {author}
  {\bibfnamefont {I.}~\bibnamefont {Tavernelli}}, \bibinfo {author}
  {\bibfnamefont {T.}~\bibnamefont {Neupert}}, \bibinfo {author} {\bibfnamefont
  {E.}~\bibnamefont {Demler}},\ and\ \bibinfo {author} {\bibfnamefont
  {Y.}~\bibnamefont {Wang}},\ }\bibfield  {title} {\bibinfo {title} {A hybrid
  quantum-classical method for electron-phonon systems},\ }\href
  {https://doi.org/10.1038/s42005-023-01353-3} {\bibfield  {journal} {\bibinfo
  {journal} {Commun. Phys.}\ }\textbf {\bibinfo {volume} {6}},\ \bibinfo
  {pages} {233} (\bibinfo {year} {2023})}\BibitemShut {NoStop}%
\bibitem [{\citenamefont {Benavides-Riveros}\ \emph {et~al.}(2022)\citenamefont
  {Benavides-Riveros}, \citenamefont {Chen}, \citenamefont {Schilling},
  \citenamefont {Mantilla},\ and\ \citenamefont
  {Pittalis}}]{PhysRevLett.129.066401}%
  \BibitemOpen
  \bibfield  {author} {\bibinfo {author} {\bibfnamefont {C.~L.}\ \bibnamefont
  {Benavides-Riveros}}, \bibinfo {author} {\bibfnamefont {L.}~\bibnamefont
  {Chen}}, \bibinfo {author} {\bibfnamefont {C.}~\bibnamefont {Schilling}},
  \bibinfo {author} {\bibfnamefont {S.}~\bibnamefont {Mantilla}},\ and\
  \bibinfo {author} {\bibfnamefont {S.}~\bibnamefont {Pittalis}},\ }\bibfield
  {title} {\bibinfo {title} {{Excitations of Quantum Many-Body Systems via
  Purified Ensembles: A Unitary-Coupled-Cluster-Based Approach}},\ }\href
  {https://doi.org/10.1103/PhysRevLett.129.066401} {\bibfield  {journal}
  {\bibinfo  {journal} {Phys. Rev. Lett.}\ }\textbf {\bibinfo {volume} {129}},\
  \bibinfo {pages} {066401} (\bibinfo {year} {2022})}\BibitemShut {NoStop}%
\bibitem [{\citenamefont {Grimsley}\ \emph {et~al.}(2020)\citenamefont
  {Grimsley}, \citenamefont {Claudino}, \citenamefont {Economou}, \citenamefont
  {Barnes},\ and\ \citenamefont {Mayhall}}]{doi:10.1021/acs.jctc.9b01083}%
  \BibitemOpen
  \bibfield  {author} {\bibinfo {author} {\bibfnamefont {H.~R.}\ \bibnamefont
  {Grimsley}}, \bibinfo {author} {\bibfnamefont {D.}~\bibnamefont {Claudino}},
  \bibinfo {author} {\bibfnamefont {S.~E.}\ \bibnamefont {Economou}}, \bibinfo
  {author} {\bibfnamefont {E.}~\bibnamefont {Barnes}},\ and\ \bibinfo {author}
  {\bibfnamefont {N.~J.}\ \bibnamefont {Mayhall}},\ }\bibfield  {title}
  {\bibinfo {title} {{Is the Trotterized UCCSD Ansatz Chemically
  Well-Defined?}},\ }\href {https://doi.org/10.1021/acs.jctc.9b01083}
  {\bibfield  {journal} {\bibinfo  {journal} {J. Chem. Theory Comput.}\
  }\textbf {\bibinfo {volume} {16}},\ \bibinfo {pages} {1} (\bibinfo {year}
  {2020})}\BibitemShut {NoStop}%
\bibitem [{\citenamefont
  {Mazziotti}(1998{\natexlab{a}})}]{mazziottiContractedSchrodingerEquation1998}%
  \BibitemOpen
  \bibfield  {author} {\bibinfo {author} {\bibfnamefont {D.~A.}\ \bibnamefont
  {Mazziotti}},\ }\bibfield  {title} {\bibinfo {title} {Contracted
  {{Schr\"odinger}} equation: {{Determining}} quantum energies and two-particle
  density matrices without wave functions},\ }\href
  {https://doi.org/10.1103/PhysRevA.57.4219} {\bibfield  {journal} {\bibinfo
  {journal} {Phys. Rev. A}\ }\textbf {\bibinfo {volume} {57}},\ \bibinfo
  {pages} {4219} (\bibinfo {year} {1998}{\natexlab{a}})}\BibitemShut {NoStop}%
\bibitem [{\citenamefont {Colmenero}\ and\ \citenamefont
  {Valdemoro}(1993)}]{Colmenero.1993}%
  \BibitemOpen
  \bibfield  {author} {\bibinfo {author} {\bibfnamefont {F.}~\bibnamefont
  {Colmenero}}\ and\ \bibinfo {author} {\bibfnamefont {C.}~\bibnamefont
  {Valdemoro}},\ }\bibfield  {title} {\bibinfo {title} {{Approximating q-order
  reduced density matrices in terms of the lower-order ones. II.
  Applications}},\ }\href {https://doi.org/10.1103/physreva.47.979} {\bibfield
  {journal} {\bibinfo  {journal} {Phys. Rev. A}\ }\textbf {\bibinfo {volume}
  {47}},\ \bibinfo {pages} {979} (\bibinfo {year} {1993})}\BibitemShut
  {NoStop}%
\bibitem [{\citenamefont {Nakatsuji}\ and\ \citenamefont
  {Yasuda}(1996)}]{Nakatsuji.1996}%
  \BibitemOpen
  \bibfield  {author} {\bibinfo {author} {\bibfnamefont {H.}~\bibnamefont
  {Nakatsuji}}\ and\ \bibinfo {author} {\bibfnamefont {K.}~\bibnamefont
  {Yasuda}},\ }\bibfield  {title} {\bibinfo {title} {{Direct Determination of
  the Quantum-Mechanical Density Matrix Using the Density Equation}},\ }\href
  {https://doi.org/10.1103/physrevlett.76.1039} {\bibfield  {journal} {\bibinfo
   {journal} {Phys. Rev. Lett.}\ }\textbf {\bibinfo {volume} {76}},\ \bibinfo
  {pages} {1039} (\bibinfo {year} {1996})}\BibitemShut {NoStop}%
\bibitem [{\citenamefont {Mazziotti}(1999)}]{Mazziotti.1999j9j}%
  \BibitemOpen
  \bibfield  {author} {\bibinfo {author} {\bibfnamefont {D.~A.}\ \bibnamefont
  {Mazziotti}},\ }\bibfield  {title} {\bibinfo {title} {{Comparison of
  contracted Schrödinger and coupled-cluster theories}},\ }\href
  {https://doi.org/10.1103/physreva.60.4396} {\bibfield  {journal} {\bibinfo
  {journal} {Phys. Rev. A}\ }\textbf {\bibinfo {volume} {60}},\ \bibinfo
  {pages} {4396} (\bibinfo {year} {1999})}\BibitemShut {NoStop}%
\bibitem [{\citenamefont {Mukherjee}\ and\ \citenamefont
  {Kutzelnigg}(2001)}]{Mukherjee.2001}%
  \BibitemOpen
  \bibfield  {author} {\bibinfo {author} {\bibfnamefont {D.}~\bibnamefont
  {Mukherjee}}\ and\ \bibinfo {author} {\bibfnamefont {W.}~\bibnamefont
  {Kutzelnigg}},\ }\bibfield  {title} {\bibinfo {title} {{Irreducible Brillouin
  conditions and contracted Schrödinger equations for n-electron systems. I.
  The equations satisfied by the density cumulants}},\ }\href
  {https://doi.org/10.1063/1.1337058} {\bibfield  {journal} {\bibinfo
  {journal} {J. Chem. Phys.}\ }\textbf {\bibinfo {volume} {114}},\ \bibinfo
  {pages} {2047} (\bibinfo {year} {2001})}\BibitemShut {NoStop}%
\bibitem [{\citenamefont {Yasuda}(2002)}]{Yasuda.2002}%
  \BibitemOpen
  \bibfield  {author} {\bibinfo {author} {\bibfnamefont {K.}~\bibnamefont
  {Yasuda}},\ }\bibfield  {title} {\bibinfo {title} {{Uniqueness of the
  solution of the contracted Schrödinger equation}},\ }\href
  {https://doi.org/10.1103/physreva.65.052121} {\bibfield  {journal} {\bibinfo
  {journal} {Phys. Rev. A}\ }\textbf {\bibinfo {volume} {65}},\ \bibinfo
  {pages} {052121} (\bibinfo {year} {2002})}\BibitemShut {NoStop}%
\bibitem [{\citenamefont {Mazziotti}(2002)}]{Mazziotti.2002}%
  \BibitemOpen
  \bibfield  {author} {\bibinfo {author} {\bibfnamefont {D.~A.}\ \bibnamefont
  {Mazziotti}},\ }\bibfield  {title} {\bibinfo {title} {{Variational method for
  solving the contracted Schrödinger equation through a projection of the
  N-particle power method onto the two-particle space}},\ }\href
  {https://doi.org/10.1063/1.1430257} {\bibfield  {journal} {\bibinfo
  {journal} {J. Chem. Phys.}\ }\textbf {\bibinfo {volume} {116}},\ \bibinfo
  {pages} {1239} (\bibinfo {year} {2002})}\BibitemShut {NoStop}%
\bibitem [{\citenamefont {Mazziotti}(2020)}]{Mazziotti.2020}%
  \BibitemOpen
  \bibfield  {author} {\bibinfo {author} {\bibfnamefont {D.~A.}\ \bibnamefont
  {Mazziotti}},\ }\bibfield  {title} {\bibinfo {title} {{Exact two-body
  expansion of the many-particle wave function}},\ }\href
  {https://doi.org/10.1103/physreva.102.030802} {\bibfield  {journal} {\bibinfo
   {journal} {Phys. Rev. A}\ }\textbf {\bibinfo {volume} {102}},\ \bibinfo
  {pages} {030802} (\bibinfo {year} {2020})},\ \Eprint
  {https://arxiv.org/abs/2010.02191} {2010.02191} \BibitemShut {NoStop}%
\bibitem [{\citenamefont {Smart}\ and\ \citenamefont
  {Mazziotti}(2024)}]{Smart.2024}%
  \BibitemOpen
  \bibfield  {author} {\bibinfo {author} {\bibfnamefont {S.~E.}\ \bibnamefont
  {Smart}}\ and\ \bibinfo {author} {\bibfnamefont {D.~A.}\ \bibnamefont
  {Mazziotti}},\ }\bibfield  {title} {\bibinfo {title} {{Verifiably exact
  solution of the electronic Schrödinger equation on quantum devices}},\
  }\href {https://doi.org/10.1103/physreva.109.022802} {\bibfield  {journal}
  {\bibinfo  {journal} {Phys. Rev. A}\ }\textbf {\bibinfo {volume} {109}},\
  \bibinfo {pages} {022802} (\bibinfo {year} {2024})}\BibitemShut {NoStop}%
\bibitem [{\citenamefont {Cohen}\ and\ \citenamefont
  {Frishberg}(1976)}]{cohenHierarchyEquationsReduced1976}%
  \BibitemOpen
  \bibfield  {author} {\bibinfo {author} {\bibfnamefont {L.}~\bibnamefont
  {Cohen}}\ and\ \bibinfo {author} {\bibfnamefont {C.}~\bibnamefont
  {Frishberg}},\ }\bibfield  {title} {\bibinfo {title} {{Hierarchy Equations
  for Reduced Density Matrices}},\ }\href
  {https://doi.org/10.1103/PhysRevA.13.927} {\bibfield  {journal} {\bibinfo
  {journal} {Phys. Rev. A}\ }\textbf {\bibinfo {volume} {13}},\ \bibinfo
  {pages} {927} (\bibinfo {year} {1976})}\BibitemShut {NoStop}%
\bibitem [{\citenamefont {Nakatsuji}(1976)}]{PhysRevA.14.41}%
  \BibitemOpen
  \bibfield  {author} {\bibinfo {author} {\bibfnamefont {H.}~\bibnamefont
  {Nakatsuji}},\ }\bibfield  {title} {\bibinfo {title} {Equation for the direct
  determination of the density matrix},\ }\href
  {https://doi.org/10.1103/PhysRevA.14.41} {\bibfield  {journal} {\bibinfo
  {journal} {Phys. Rev. A}\ }\textbf {\bibinfo {volume} {14}},\ \bibinfo
  {pages} {41} (\bibinfo {year} {1976})}\BibitemShut {NoStop}%
\bibitem [{\citenamefont {Valdemoro}\ \emph {et~al.}(2007)\citenamefont
  {Valdemoro}, \citenamefont {Tel}, \citenamefont {Alcoba},\ and\ \citenamefont
  {P{\'e}rez-Rome\-ro}}]{Valdemoro2007}%
  \BibitemOpen
  \bibfield  {author} {\bibinfo {author} {\bibfnamefont {C.}~\bibnamefont
  {Valdemoro}}, \bibinfo {author} {\bibfnamefont {L.}~\bibnamefont {Tel}},
  \bibinfo {author} {\bibfnamefont {D.}~\bibnamefont {Alcoba}},\ and\ \bibinfo
  {author} {\bibfnamefont {E.}~\bibnamefont {P{\'e}rez-Rome\-ro}},\ }\bibfield
  {title} {\bibinfo {title} {{The contracted Schr{\"o}dinger equation
  methodology: study of the third-order correlation effects}},\ }\href
  {https://doi.org/10.1007/s00214-007-0337-z} {\bibfield  {journal} {\bibinfo
  {journal} {Theor. Chem. Account.}\ }\textbf {\bibinfo {volume} {118}},\
  \bibinfo {pages} {503} (\bibinfo {year} {2007})}\BibitemShut {NoStop}%
\bibitem [{\citenamefont {Mazziotti}(2006{\natexlab{a}})}]{Mazziotti.20060v3}%
  \BibitemOpen
  \bibfield  {author} {\bibinfo {author} {\bibfnamefont {D.~A.}\ \bibnamefont
  {Mazziotti}},\ }\bibfield  {title} {\bibinfo {title} {{Anti-Hermitian
  Contracted Schrödinger Equation: Direct Determination of the Two-Electron
  Reduced Density Matrices of Many-Electron Molecules}},\ }\href
  {https://doi.org/10.1103/physrevlett.97.143002} {\bibfield  {journal}
  {\bibinfo  {journal} {Phys. Rev. Lett.}\ }\textbf {\bibinfo {volume} {97}},\
  \bibinfo {pages} {143002} (\bibinfo {year} {2006}{\natexlab{a}})}\BibitemShut
  {NoStop}%
\bibitem [{\citenamefont {Mazziotti}(2007{\natexlab{a}})}]{Mazziotti.2007}%
  \BibitemOpen
  \bibfield  {author} {\bibinfo {author} {\bibfnamefont {D.~A.}\ \bibnamefont
  {Mazziotti}},\ }\bibfield  {title} {\bibinfo {title} {{Anti-Hermitian part of
  the contracted Schrödinger equation for the direct calculation of
  two-electron reduced density matrices}},\ }\href
  {https://doi.org/10.1103/physreva.75.022505} {\bibfield  {journal} {\bibinfo
  {journal} {Phys. Rev. A}\ }\textbf {\bibinfo {volume} {75}},\ \bibinfo
  {pages} {022505} (\bibinfo {year} {2007}{\natexlab{a}})}\BibitemShut
  {NoStop}%
\bibitem [{\citenamefont {Mazziotti}(2007{\natexlab{b}})}]{Mazziotti.2007k2h}%
  \BibitemOpen
  \bibfield  {author} {\bibinfo {author} {\bibfnamefont {D.~A.}\ \bibnamefont
  {Mazziotti}},\ }\bibfield  {title} {\bibinfo {title} {{Multireference
  many-electron correlation energies from two-electron reduced density matrices
  computed by solving the anti-Hermitian contracted Schrödinger equation}},\
  }\href {https://doi.org/10.1103/physreva.76.052502} {\bibfield  {journal}
  {\bibinfo  {journal} {Phys. Rev. A}\ }\textbf {\bibinfo {volume} {76}},\
  \bibinfo {pages} {052502} (\bibinfo {year} {2007}{\natexlab{b}})}\BibitemShut
  {NoStop}%
\bibitem [{\citenamefont {Snyder}\ and\ \citenamefont
  {Mazziotti}(2011)}]{Snyder.2011u3}%
  \BibitemOpen
  \bibfield  {author} {\bibinfo {author} {\bibfnamefont {J.~W.}\ \bibnamefont
  {Snyder}}\ and\ \bibinfo {author} {\bibfnamefont {D.~A.}\ \bibnamefont
  {Mazziotti}},\ }\bibfield  {title} {\bibinfo {title} {{Photoexcited
  conversion of gauche-1,3-butadiene to bicyclobutane via a conical
  intersection: Energies and reduced density matrices from the anti-Hermitian
  contracted Schrödinger equation}},\ }\href
  {https://doi.org/10.1063/1.3606466} {\bibfield  {journal} {\bibinfo
  {journal} {J. Chem. Phys.}\ }\textbf {\bibinfo {volume} {135}},\ \bibinfo
  {pages} {024107} (\bibinfo {year} {2011})}\BibitemShut {NoStop}%
\bibitem [{\citenamefont {Gidofalvi}\ and\ \citenamefont
  {Mazziotti}(2009)}]{Gidofalvi.2009}%
  \BibitemOpen
  \bibfield  {author} {\bibinfo {author} {\bibfnamefont {G.}~\bibnamefont
  {Gidofalvi}}\ and\ \bibinfo {author} {\bibfnamefont {D.~A.}\ \bibnamefont
  {Mazziotti}},\ }\bibfield  {title} {\bibinfo {title} {{Direct calculation of
  excited-state electronic energies and two-electron reduced density matrices
  from the anti-Hermitian contracted Schrödinger equation}},\ }\href
  {https://doi.org/10.1103/physreva.80.022507} {\bibfield  {journal} {\bibinfo
  {journal} {Phys. Rev. A}\ }\textbf {\bibinfo {volume} {80}},\ \bibinfo
  {pages} {022507} (\bibinfo {year} {2009})}\BibitemShut {NoStop}%
\bibitem [{\citenamefont {Alcoba}\ \emph {et~al.}(2011)\citenamefont {Alcoba},
  \citenamefont {Valdemoro}, \citenamefont {Tel}, \citenamefont
  {Pe{\'e}rez-Romero},\ and\ \citenamefont {O{\~{n}}a}}]{Alcoba.2011}%
  \BibitemOpen
  \bibfield  {author} {\bibinfo {author} {\bibfnamefont {D.~R.}\ \bibnamefont
  {Alcoba}}, \bibinfo {author} {\bibfnamefont {C.}~\bibnamefont {Valdemoro}},
  \bibinfo {author} {\bibfnamefont {L.~M.}\ \bibnamefont {Tel}}, \bibinfo
  {author} {\bibfnamefont {E.}~\bibnamefont {Pe{\'e}rez-Romero}},\ and\
  \bibinfo {author} {\bibfnamefont {O.~B.}\ \bibnamefont {O{\~{n}}a}},\
  }\bibfield  {title} {\bibinfo {title} {{Optimized Solution Procedure of the
  G-Particle−Hole Hypervirial Equation for Multiplets: Application to Doublet
  and Triplet States}},\ }\href {https://doi.org/10.1021/jp109018t} {\bibfield
  {journal} {\bibinfo  {journal} {J. Phys. Chem. A}\ }\textbf {\bibinfo
  {volume} {115}},\ \bibinfo {pages} {2599} (\bibinfo {year}
  {2011})}\BibitemShut {NoStop}%
\bibitem [{\citenamefont {Boyn}\ and\ \citenamefont
  {Mazziotti}(2021)}]{Boyn.2021}%
  \BibitemOpen
  \bibfield  {author} {\bibinfo {author} {\bibfnamefont {J.-N.}\ \bibnamefont
  {Boyn}}\ and\ \bibinfo {author} {\bibfnamefont {D.~A.}\ \bibnamefont
  {Mazziotti}},\ }\bibfield  {title} {\bibinfo {title} {{Accurate
  singlet–triplet gaps in biradicals via the spin averaged anti-Hermitian
  contracted Schrödinger equation}},\ }\href
  {https://doi.org/10.1063/5.0045007} {\bibfield  {journal} {\bibinfo
  {journal} {J. Chem. Phys.}\ }\textbf {\bibinfo {volume} {154}},\ \bibinfo
  {pages} {134103} (\bibinfo {year} {2021})},\ \Eprint
  {https://arxiv.org/abs/2104.00626} {2104.00626} \BibitemShut {NoStop}%
\bibitem [{\citenamefont
  {Mazziotti}(2006{\natexlab{b}})}]{mazziottiQuantumChemistryWave2006}%
  \BibitemOpen
  \bibfield  {author} {\bibinfo {author} {\bibfnamefont {D.~A.}\ \bibnamefont
  {Mazziotti}},\ }\bibfield  {title} {\bibinfo {title} {Quantum {{Chemistry}}
  without {{Wave Functions}}:\, {{Two-Electron Reduced Density Matrices}}},\
  }\href {https://doi.org/10.1021/ar050029d} {\bibfield  {journal} {\bibinfo
  {journal} {Acc. Chem. Res.}\ }\textbf {\bibinfo {volume} {39}},\ \bibinfo
  {pages} {207} (\bibinfo {year} {2006}{\natexlab{b}})}\BibitemShut {NoStop}%
\bibitem [{\citenamefont {Mazziotti}(2007{\natexlab{c}})}]{ch8}%
  \BibitemOpen
  \bibfield  {author} {\bibinfo {author} {\bibfnamefont {D.~A.}\ \bibnamefont
  {Mazziotti}},\ }\bibinfo {title} {{Contracted Schrödinger Equation}},\ in\
  \href {https://doi.org/https://doi.org/10.1002/9780470106600.ch8} {\emph
  {\bibinfo {booktitle} {{Reduced‐Density‐Matrix Mechanics: With
  Application to Many‐Electron Atoms and Molecules}}}}\ (\bibinfo
  {publisher} {John Wiley \& Sons, Ltd},\ \bibinfo {year} {2007})\
  Chap.~\bibinfo {chapter} {8}, pp.\ \bibinfo {pages} {165--203}\BibitemShut
  {NoStop}%
\bibitem [{\citenamefont
  {Mazziotti}(2012)}]{mazziottiTwoElectronReducedDensity2012}%
  \BibitemOpen
  \bibfield  {author} {\bibinfo {author} {\bibfnamefont {D.~A.}\ \bibnamefont
  {Mazziotti}},\ }\bibfield  {title} {\bibinfo {title} {Two-{{Electron Reduced
  Density Matrix}} as the {{Basic Variable}} in {{Many-Electron Quantum
  Chemistry}} and {{Physics}}},\ }\href {https://doi.org/10.1021/cr2000493}
  {\bibfield  {journal} {\bibinfo  {journal} {Chem. Rev.}\ }\textbf {\bibinfo
  {volume} {112}},\ \bibinfo {pages} {244} (\bibinfo {year}
  {2012})}\BibitemShut {NoStop}%
\bibitem [{\citenamefont {Mazziotti}(1998{\natexlab{b}})}]{PhysRevA.57.4219}%
  \BibitemOpen
  \bibfield  {author} {\bibinfo {author} {\bibfnamefont {D.~A.}\ \bibnamefont
  {Mazziotti}},\ }\bibfield  {title} {\bibinfo {title} {{Contracted
  Schr\"odinger equation: Determining quantum energies and two-particle density
  matrices without wave functions}},\ }\href
  {https://doi.org/10.1103/PhysRevA.57.4219} {\bibfield  {journal} {\bibinfo
  {journal} {Phys. Rev. A}\ }\textbf {\bibinfo {volume} {57}},\ \bibinfo
  {pages} {4219} (\bibinfo {year} {1998}{\natexlab{b}})}\BibitemShut {NoStop}%
\bibitem [{\citenamefont {Rosina}(1968)}]{rosina}%
  \BibitemOpen
  \bibfield  {author} {\bibinfo {author} {\bibfnamefont {M.}~\bibnamefont
  {Rosina}},\ }in\ \href {https://books.google.it/books?id=Ln4gAQAAIAAJ} {\emph
  {\bibinfo {booktitle} {Reduced Density Matrices with Applications to Physical
  and Chemical Systems}}},\ \bibinfo {editor} {edited by\ \bibinfo {editor}
  {\bibfnamefont {A.}~\bibnamefont {Coleman}}\ and\ \bibinfo {editor}
  {\bibfnamefont {R.}~\bibnamefont {Erdahl}}}\ (\bibinfo  {publisher} {Queen's
  papers in pure and applied mathematics},\ \bibinfo {year} {1968})\BibitemShut
  {NoStop}%
\bibitem [{\citenamefont {Mazziotti}(2004)}]{PhysRevA.69.012507}%
  \BibitemOpen
  \bibfield  {author} {\bibinfo {author} {\bibfnamefont {D.~A.}\ \bibnamefont
  {Mazziotti}},\ }\bibfield  {title} {\bibinfo {title} {Exactness of wave
  functions from two-body exponential transformations in many-body quantum
  theory},\ }\href {https://doi.org/10.1103/PhysRevA.69.012507} {\bibfield
  {journal} {\bibinfo  {journal} {Phys. Rev. A}\ }\textbf {\bibinfo {volume}
  {69}},\ \bibinfo {pages} {012507} (\bibinfo {year} {2004})}\BibitemShut
  {NoStop}%
\bibitem [{\citenamefont {Smart}\ and\ \citenamefont
  {Mazziotti}(2021)}]{PhysRevLett.126.070504}%
  \BibitemOpen
  \bibfield  {author} {\bibinfo {author} {\bibfnamefont {S.~E.}\ \bibnamefont
  {Smart}}\ and\ \bibinfo {author} {\bibfnamefont {D.~A.}\ \bibnamefont
  {Mazziotti}},\ }\bibfield  {title} {\bibinfo {title} {{Quantum Solver of
  Contracted Eigenvalue Equations for Scalable Molecular Simulations on Quantum
  Computing Devices}},\ }\href {https://doi.org/10.1103/PhysRevLett.126.070504}
  {\bibfield  {journal} {\bibinfo  {journal} {Phys. Rev. Lett.}\ }\textbf
  {\bibinfo {volume} {126}},\ \bibinfo {pages} {070504} (\bibinfo {year}
  {2021})}\BibitemShut {NoStop}%
\bibitem [{\citenamefont {Boyn}\ \emph {et~al.}(2021)\citenamefont {Boyn},
  \citenamefont {Lykhin}, \citenamefont {Smart}, \citenamefont {Gagliardi},\
  and\ \citenamefont {Mazziotti}}]{Boyn.2021u94}%
  \BibitemOpen
  \bibfield  {author} {\bibinfo {author} {\bibfnamefont {J.-N.}\ \bibnamefont
  {Boyn}}, \bibinfo {author} {\bibfnamefont {A.~O.}\ \bibnamefont {Lykhin}},
  \bibinfo {author} {\bibfnamefont {S.~E.}\ \bibnamefont {Smart}}, \bibinfo
  {author} {\bibfnamefont {L.}~\bibnamefont {Gagliardi}},\ and\ \bibinfo
  {author} {\bibfnamefont {D.~A.}\ \bibnamefont {Mazziotti}},\ }\bibfield
  {title} {\bibinfo {title} {{Quantum-classical hybrid algorithm for the
  simulation of all-electron correlation}},\ }\href
  {https://doi.org/10.1063/5.0074842} {\bibfield  {journal} {\bibinfo
  {journal} {J. Chem. Phys.}\ }\textbf {\bibinfo {volume} {155}},\ \bibinfo
  {pages} {244106} (\bibinfo {year} {2021})},\ \Eprint
  {https://arxiv.org/abs/2106.11972} {2106.11972} \BibitemShut {NoStop}%
\bibitem [{\citenamefont {Smart}\ \emph {et~al.}(2022)\citenamefont {Smart},
  \citenamefont {Boyn},\ and\ \citenamefont {Mazziotti}}]{Smart.2022w8u}%
  \BibitemOpen
  \bibfield  {author} {\bibinfo {author} {\bibfnamefont {S.~E.}\ \bibnamefont
  {Smart}}, \bibinfo {author} {\bibfnamefont {J.-N.}\ \bibnamefont {Boyn}},\
  and\ \bibinfo {author} {\bibfnamefont {D.~A.}\ \bibnamefont {Mazziotti}},\
  }\bibfield  {title} {\bibinfo {title} {{Resolving correlated states of
  benzyne with an error-mitigated contracted quantum eigensolver}},\ }\href
  {https://doi.org/10.1103/physreva.105.022405} {\bibfield  {journal} {\bibinfo
   {journal} {Phys. Rev. A}\ }\textbf {\bibinfo {volume} {105}},\ \bibinfo
  {pages} {022405} (\bibinfo {year} {2022})},\ \Eprint
  {https://arxiv.org/abs/2103.06876} {2103.06876} \BibitemShut {NoStop}%
\bibitem [{\citenamefont {Smart}\ and\ \citenamefont
  {Mazziotti}(2022)}]{Smart.2022}%
  \BibitemOpen
  \bibfield  {author} {\bibinfo {author} {\bibfnamefont {S.~E.}\ \bibnamefont
  {Smart}}\ and\ \bibinfo {author} {\bibfnamefont {D.~A.}\ \bibnamefont
  {Mazziotti}},\ }\bibfield  {title} {\bibinfo {title} {{Many-fermion
  simulation from the contracted quantum eigensolver without fermionic encoding
  of the wave function}},\ }\href {https://doi.org/10.1103/physreva.105.062424}
  {\bibfield  {journal} {\bibinfo  {journal} {Phys. Rev. A}\ }\textbf {\bibinfo
  {volume} {105}},\ \bibinfo {pages} {062424} (\bibinfo {year} {2022})},\
  \Eprint {https://arxiv.org/abs/2205.01725} {2205.01725} \BibitemShut
  {NoStop}%
\bibitem [{\citenamefont {Wang}\ and\ \citenamefont
  {Mazziotti}(2023)}]{Wang.20232b}%
  \BibitemOpen
  \bibfield  {author} {\bibinfo {author} {\bibfnamefont {Y.}~\bibnamefont
  {Wang}}\ and\ \bibinfo {author} {\bibfnamefont {D.~A.}\ \bibnamefont
  {Mazziotti}},\ }\bibfield  {title} {\bibinfo {title} {{Electronic excited
  states from a variance-based contracted quantum eigensolver}},\ }\href
  {https://doi.org/10.1103/physreva.108.022814} {\bibfield  {journal} {\bibinfo
   {journal} {Phys. Rev. A}\ }\textbf {\bibinfo {volume} {108}},\ \bibinfo
  {pages} {022814} (\bibinfo {year} {2023})}\BibitemShut {NoStop}%
\bibitem [{\citenamefont {Wang}\ \emph
  {et~al.}(2023{\natexlab{a}})\citenamefont {Wang}, \citenamefont {Smith},\
  and\ \citenamefont {Mazziotti}}]{wang2023boson}%
  \BibitemOpen
  \bibfield  {author} {\bibinfo {author} {\bibfnamefont {Y.}~\bibnamefont
  {Wang}}, \bibinfo {author} {\bibfnamefont {L.~M.}\ \bibnamefont {Smith}},\
  and\ \bibinfo {author} {\bibfnamefont {D.~A.}\ \bibnamefont {Mazziotti}},\
  }\bibfield  {title} {\bibinfo {title} {Quantum simulation of bosons with the
  contracted quantum eigensolver},\ }\href
  {https://iopscience.iop.org/article/10.1088/1367-2630/acf9c3} {\bibfield
  {journal} {\bibinfo  {journal} {New J. Phys.}\ }\textbf {\bibinfo {volume}
  {25}},\ \bibinfo {pages} {103005} (\bibinfo {year}
  {2023}{\natexlab{a}})}\BibitemShut {NoStop}%
\bibitem [{\citenamefont {{Lang}}\ and\ \citenamefont {{Firsov}}(1963)}]{Lang}%
  \BibitemOpen
  \bibfield  {author} {\bibinfo {author} {\bibfnamefont {I.~G.}\ \bibnamefont
  {{Lang}}}\ and\ \bibinfo {author} {\bibfnamefont {Y.~A.}\ \bibnamefont
  {{Firsov}}},\ }\bibfield  {title} {\bibinfo {title} {{Kinetic Theory of
  Semiconductors with Low Mobility}},\ }\href
  {http://www.jetp.ras.ru/cgi-bin/e/index/e/16/5/p1301?a=list} {\bibfield
  {journal} {\bibinfo  {journal} {Sov. Phys. JETP}\ }\textbf {\bibinfo {volume}
  {16}},\ \bibinfo {pages} {1301} (\bibinfo {year} {1963})}\BibitemShut
  {NoStop}%
\bibitem [{\citenamefont {Schrieffer}\ and\ \citenamefont
  {Wolff}(1966)}]{PhysRev.149.491}%
  \BibitemOpen
  \bibfield  {author} {\bibinfo {author} {\bibfnamefont {J.}~\bibnamefont
  {Schrieffer}}\ and\ \bibinfo {author} {\bibfnamefont {P.}~\bibnamefont
  {Wolff}},\ }\bibfield  {title} {\bibinfo {title} {{Relation between the
  Anderson and Kondo Hamiltonians}},\ }\href
  {https://doi.org/10.1103/PhysRev.149.491} {\bibfield  {journal} {\bibinfo
  {journal} {Phys. Rev.}\ }\textbf {\bibinfo {volume} {149}},\ \bibinfo {pages}
  {491} (\bibinfo {year} {1966})}\BibitemShut {NoStop}%
\bibitem [{\citenamefont {Hu}\ \emph {et~al.}(2020)\citenamefont {Hu},
  \citenamefont {Xia},\ and\ \citenamefont {Kais}}]{Hu2020}%
  \BibitemOpen
  \bibfield  {author} {\bibinfo {author} {\bibfnamefont {Z.}~\bibnamefont
  {Hu}}, \bibinfo {author} {\bibfnamefont {R.}~\bibnamefont {Xia}},\ and\
  \bibinfo {author} {\bibfnamefont {S.}~\bibnamefont {Kais}},\ }\bibfield
  {title} {\bibinfo {title} {A quantum algorithm for evolving open quantum
  dynamics on quantum computing devices},\ }\href
  {https://doi.org/10.1038/s41598-020-60321-x} {\bibfield  {journal} {\bibinfo
  {journal} {Sci. Rep.}\ }\textbf {\bibinfo {volume} {10}},\ \bibinfo {pages}
  {3301} (\bibinfo {year} {2020})}\BibitemShut {NoStop}%
\bibitem [{\citenamefont {Wang}\ \emph
  {et~al.}(2023{\natexlab{b}})\citenamefont {Wang}, \citenamefont {Mulvihill},
  \citenamefont {Hu}, \citenamefont {Lyu}, \citenamefont {Shivpuje},
  \citenamefont {Liu}, \citenamefont {Soley}, \citenamefont {Geva},
  \citenamefont {Batista},\ and\ \citenamefont
  {Kais}}]{doi:10.1021/acs.jctc.3c00316}%
  \BibitemOpen
  \bibfield  {author} {\bibinfo {author} {\bibfnamefont {Y.}~\bibnamefont
  {Wang}}, \bibinfo {author} {\bibfnamefont {E.}~\bibnamefont {Mulvihill}},
  \bibinfo {author} {\bibfnamefont {Z.}~\bibnamefont {Hu}}, \bibinfo {author}
  {\bibfnamefont {N.}~\bibnamefont {Lyu}}, \bibinfo {author} {\bibfnamefont
  {S.}~\bibnamefont {Shivpuje}}, \bibinfo {author} {\bibfnamefont
  {Y.}~\bibnamefont {Liu}}, \bibinfo {author} {\bibfnamefont {M.}~\bibnamefont
  {Soley}}, \bibinfo {author} {\bibfnamefont {E.}~\bibnamefont {Geva}},
  \bibinfo {author} {\bibfnamefont {V.}~\bibnamefont {Batista}},\ and\ \bibinfo
  {author} {\bibfnamefont {S.}~\bibnamefont {Kais}},\ }\bibfield  {title}
  {\bibinfo {title} {{Simulating Open Quantum System Dynamics on NISQ Computers
  with Generalized Quantum Master Equations}},\ }\href
  {https://doi.org/10.1021/acs.jctc.3c00316} {\bibfield  {journal} {\bibinfo
  {journal} {J. Chem. Theory Comput.}\ }\textbf {\bibinfo {volume} {19}},\
  \bibinfo {pages} {4851} (\bibinfo {year} {2023}{\natexlab{b}})}\BibitemShut
  {NoStop}%
\bibitem [{\citenamefont {Motta}\ \emph {et~al.}(2020)\citenamefont {Motta},
  \citenamefont {Sun}, \citenamefont {Tan}, \citenamefont {O'Rourke},
  \citenamefont {Ye}, \citenamefont {Minnich}, \citenamefont {Brand{\~a}o},\
  and\ \citenamefont {Chan}}]{Motta2020}%
  \BibitemOpen
  \bibfield  {author} {\bibinfo {author} {\bibfnamefont {M.}~\bibnamefont
  {Motta}}, \bibinfo {author} {\bibfnamefont {C.}~\bibnamefont {Sun}}, \bibinfo
  {author} {\bibfnamefont {A.}~\bibnamefont {Tan}}, \bibinfo {author}
  {\bibfnamefont {M.}~\bibnamefont {O'Rourke}}, \bibinfo {author}
  {\bibfnamefont {E.}~\bibnamefont {Ye}}, \bibinfo {author} {\bibfnamefont
  {A.}~\bibnamefont {Minnich}}, \bibinfo {author} {\bibfnamefont
  {F.}~\bibnamefont {Brand{\~a}o}},\ and\ \bibinfo {author} {\bibfnamefont
  {G.}~\bibnamefont {Chan}},\ }\bibfield  {title} {\bibinfo {title}
  {Determining eigenstates and thermal states on a quantum computer using
  quantum imaginary time evolution},\ }\href
  {https://doi.org/10.1038/s41567-019-0704-4} {\bibfield  {journal} {\bibinfo
  {journal} {Nat. Phys.}\ }\textbf {\bibinfo {volume} {16}},\ \bibinfo {pages}
  {205} (\bibinfo {year} {2020})}\BibitemShut {NoStop}%
\bibitem [{\citenamefont {McArdle}\ \emph {et~al.}(2019)\citenamefont
  {McArdle}, \citenamefont {Jones}, \citenamefont {Endo}, \citenamefont {Li},
  \citenamefont {Benjamin},\ and\ \citenamefont {Yuan}}]{McArdle2019}%
  \BibitemOpen
  \bibfield  {author} {\bibinfo {author} {\bibfnamefont {S.}~\bibnamefont
  {McArdle}}, \bibinfo {author} {\bibfnamefont {T.}~\bibnamefont {Jones}},
  \bibinfo {author} {\bibfnamefont {S.}~\bibnamefont {Endo}}, \bibinfo {author}
  {\bibfnamefont {Y.}~\bibnamefont {Li}}, \bibinfo {author} {\bibfnamefont
  {S.}~\bibnamefont {Benjamin}},\ and\ \bibinfo {author} {\bibfnamefont
  {X.}~\bibnamefont {Yuan}},\ }\bibfield  {title} {\bibinfo {title}
  {Variational ansatz-based quantum simulation of imaginary time evolution},\
  }\href {https://doi.org/10.1038/s41534-019-0187-2} {\bibfield  {journal}
  {\bibinfo  {journal} {npj Quantum Inf.}\ }\textbf {\bibinfo {volume} {5}},\
  \bibinfo {pages} {75} (\bibinfo {year} {2019})}\BibitemShut {NoStop}%
\bibitem [{\citenamefont {Jaynes}\ and\ \citenamefont
  {Cummings}(1963)}]{1443594}%
  \BibitemOpen
  \bibfield  {author} {\bibinfo {author} {\bibfnamefont {E.}~\bibnamefont
  {Jaynes}}\ and\ \bibinfo {author} {\bibfnamefont {F.}~\bibnamefont
  {Cummings}},\ }\bibfield  {title} {\bibinfo {title} {Comparison of quantum
  and semiclassical radiation theories with application to the beam maser},\
  }\href {https://doi.org/10.1109/PROC.1963.1664} {\bibfield  {journal}
  {\bibinfo  {journal} {Proceedings of the IEEE}\ }\textbf {\bibinfo {volume}
  {51}},\ \bibinfo {pages} {89} (\bibinfo {year} {1963})}\BibitemShut {NoStop}%
\bibitem [{\citenamefont {Tavis}\ and\ \citenamefont
  {Cummings}(1968)}]{tavisExactSolutionMolecule1968}%
  \BibitemOpen
  \bibfield  {author} {\bibinfo {author} {\bibfnamefont {M.}~\bibnamefont
  {Tavis}}\ and\ \bibinfo {author} {\bibfnamefont {F.}~\bibnamefont
  {Cummings}},\ }\bibfield  {title} {\bibinfo {title} {{Exact {{Solution}} for
  an $N$-{{Molecule---Radiation-Field Hamiltonian}}}},\ }\href
  {https://doi.org/10.1103/PhysRev.170.379} {\bibfield  {journal} {\bibinfo
  {journal} {Phys. Rev.}\ }\textbf {\bibinfo {volume} {170}},\ \bibinfo {pages}
  {379} (\bibinfo {year} {1968})}\BibitemShut {NoStop}%
\bibitem [{\citenamefont {Tavis}\ and\ \citenamefont
  {Cummings}(1969)}]{tavisApproximateSolutionsMoleculeRadiationField1969}%
  \BibitemOpen
  \bibfield  {author} {\bibinfo {author} {\bibfnamefont {M.}~\bibnamefont
  {Tavis}}\ and\ \bibinfo {author} {\bibfnamefont {F.}~\bibnamefont
  {Cummings}},\ }\bibfield  {title} {\bibinfo {title} {{Approximate
  {{Solutions}} for an $N$-{{Molecule-Radiation-Field Hamiltonian}}}},\ }\href
  {https://doi.org/10.1103/PhysRev.188.692} {\bibfield  {journal} {\bibinfo
  {journal} {Phys. Rev.}\ }\textbf {\bibinfo {volume} {188}},\ \bibinfo {pages}
  {692} (\bibinfo {year} {1969})}\BibitemShut {NoStop}%
\bibitem [{\citenamefont {Blaha}\ \emph {et~al.}(2022)\citenamefont {Blaha},
  \citenamefont {Johnson}, \citenamefont {Rauschenbeutel},\ and\ \citenamefont
  {Volz}}]{PhysRevA.105.013719}%
  \BibitemOpen
  \bibfield  {author} {\bibinfo {author} {\bibfnamefont {M.}~\bibnamefont
  {Blaha}}, \bibinfo {author} {\bibfnamefont {A.}~\bibnamefont {Johnson}},
  \bibinfo {author} {\bibfnamefont {A.}~\bibnamefont {Rauschenbeutel}},\ and\
  \bibinfo {author} {\bibfnamefont {J.}~\bibnamefont {Volz}},\ }\bibfield
  {title} {\bibinfo {title} {{Beyond the Tavis-Cummings model: Revisiting
  cavity QED with ensembles of quantum emitters}},\ }\href
  {https://doi.org/10.1103/PhysRevA.105.013719} {\bibfield  {journal} {\bibinfo
   {journal} {Phys. Rev. A}\ }\textbf {\bibinfo {volume} {105}},\ \bibinfo
  {pages} {013719} (\bibinfo {year} {2022})}\BibitemShut {NoStop}%
\bibitem [{\citenamefont {Castaños}\ \emph {et~al.}(2009)\citenamefont
  {Castaños}, \citenamefont {Nahmad-Achar}, \citenamefont {López-Peña},\
  and\ \citenamefont {Hirsch}}]{Castanos_2009}%
  \BibitemOpen
  \bibfield  {author} {\bibinfo {author} {\bibfnamefont {O.}~\bibnamefont
  {Castaños}}, \bibinfo {author} {\bibfnamefont {E.}~\bibnamefont
  {Nahmad-Achar}}, \bibinfo {author} {\bibfnamefont {R.}~\bibnamefont
  {López-Peña}},\ and\ \bibinfo {author} {\bibfnamefont {J.}~\bibnamefont
  {Hirsch}},\ }\bibfield  {title} {\bibinfo {title} {{Analytic approximation of
  the Tavis-Cum\-mings ground state via projected states}},\ }\href
  {https://doi.org/10.1088/0031-8949/80/05/055401} {\bibfield  {journal}
  {\bibinfo  {journal} {Phys. Scr.}\ }\textbf {\bibinfo {volume} {80}},\
  \bibinfo {pages} {055401} (\bibinfo {year} {2009})}\BibitemShut {NoStop}%
\bibitem [{\citenamefont {Gera}\ and\ \citenamefont
  {Sebastian}(2022)}]{geraEffectsDisorderPolaritonic2022}%
  \BibitemOpen
  \bibfield  {author} {\bibinfo {author} {\bibfnamefont {T.}~\bibnamefont
  {Gera}}\ and\ \bibinfo {author} {\bibfnamefont {K.}~\bibnamefont
  {Sebastian}},\ }\bibfield  {title} {\bibinfo {title} {{Effects of disorder on
  polaritonic and dark states in a cavity using the disordered
  Tavis{\textendash}{{Cummings}} model}},\ }\href
  {https://doi.org/10.1063/5.0086027} {\bibfield  {journal} {\bibinfo
  {journal} {J. Chem. Phys.}\ }\textbf {\bibinfo {volume} {156}},\ \bibinfo
  {pages} {194304} (\bibinfo {year} {2022})}\BibitemShut {NoStop}%
\bibitem [{\citenamefont {De~Bernardis}\ \emph {et~al.}(2022)\citenamefont
  {De~Bernardis}, \citenamefont {Jeannin}, \citenamefont {Manceau},
  \citenamefont {Colombelli}, \citenamefont {Tredicucci},\ and\ \citenamefont
  {Carusotto}}]{PhysRevB.106.224206}%
  \BibitemOpen
  \bibfield  {author} {\bibinfo {author} {\bibfnamefont {D.}~\bibnamefont
  {De~Bernardis}}, \bibinfo {author} {\bibfnamefont {M.}~\bibnamefont
  {Jeannin}}, \bibinfo {author} {\bibfnamefont {J.-M.}\ \bibnamefont
  {Manceau}}, \bibinfo {author} {\bibfnamefont {R.}~\bibnamefont {Colombelli}},
  \bibinfo {author} {\bibfnamefont {A.}~\bibnamefont {Tredicucci}},\ and\
  \bibinfo {author} {\bibfnamefont {I.}~\bibnamefont {Carusotto}},\ }\bibfield
  {title} {\bibinfo {title} {Mag\-ne\-tic-field-induced cavity protection for
  intersubband polaritons},\ }\href
  {https://doi.org/10.1103/PhysRevB.106.224206} {\bibfield  {journal} {\bibinfo
   {journal} {Phys. Rev. B}\ }\textbf {\bibinfo {volume} {106}},\ \bibinfo
  {pages} {224206} (\bibinfo {year} {2022})}\BibitemShut {NoStop}%
\bibitem [{\citenamefont {Todorov}\ and\ \citenamefont
  {Sirtori}(2012)}]{PhysRevB.85.045304}%
  \BibitemOpen
  \bibfield  {author} {\bibinfo {author} {\bibfnamefont {Y.}~\bibnamefont
  {Todorov}}\ and\ \bibinfo {author} {\bibfnamefont {C.}~\bibnamefont
  {Sirtori}},\ }\bibfield  {title} {\bibinfo {title} {Intersubband polaritons
  in the electrical dipole gauge},\ }\href
  {https://doi.org/10.1103/PhysRevB.85.045304} {\bibfield  {journal} {\bibinfo
  {journal} {Phys. Rev. B}\ }\textbf {\bibinfo {volume} {85}},\ \bibinfo
  {pages} {045304} (\bibinfo {year} {2012})}\BibitemShut {NoStop}%
\bibitem [{\citenamefont {K{\"o}hn}\ and\ \citenamefont
  {Tajti}(2007)}]{kohnCanCoupledclusterTheory2007}%
  \BibitemOpen
  \bibfield  {author} {\bibinfo {author} {\bibfnamefont {A.}~\bibnamefont
  {K{\"o}hn}}\ and\ \bibinfo {author} {\bibfnamefont {A.}~\bibnamefont
  {Tajti}},\ }\bibfield  {title} {\bibinfo {title} {Can coupled-cluster theory
  treat conical intersections?},\ }\href {https://doi.org/10.1063/1.2755681}
  {\bibfield  {journal} {\bibinfo  {journal} {J. Chem. Phys.}\ }\textbf
  {\bibinfo {volume} {127}},\ \bibinfo {pages} {044105} (\bibinfo {year}
  {2007})}\BibitemShut {NoStop}%
\bibitem [{\citenamefont {Benavides-Riveros}\ \emph {et~al.}(2023)\citenamefont
  {Benavides-Riveros}, \citenamefont {Wang}, \citenamefont {Warren},\ and\
  \citenamefont {Mazziotti}}]{benavidesriveros2023quantum}%
  \BibitemOpen
  \bibfield  {author} {\bibinfo {author} {\bibfnamefont {C.~L.}\ \bibnamefont
  {Benavides-Riveros}}, \bibinfo {author} {\bibfnamefont {Y.}~\bibnamefont
  {Wang}}, \bibinfo {author} {\bibfnamefont {S.}~\bibnamefont {Warren}},\ and\
  \bibinfo {author} {\bibfnamefont {D.~A.}\ \bibnamefont {Mazziotti}},\
  }\href@noop {} {\bibinfo {title} {Quantum simulation of excited states from
  parallel contracted quantum eigensolvers}} (\bibinfo {year} {2023}),\ \Eprint
  {https://arxiv.org/abs/2311.05058} {arXiv:2311.05058 [quant-ph]} \BibitemShut
  {NoStop}%
\end{thebibliography}%
\end{document}